\documentclass[fleqn,usenatbib]{mnras}

\usepackage{mathptmx}

\usepackage[T1]{fontenc}
\usepackage{soul}

\DeclareRobustCommand{\VAN}[3]{#2}
\let\VANthebibliography\thebibliography
\def\thebibliography{\DeclareRobustCommand{\VAN}[3]{##3}\VANthebibliography}

\usepackage{graphicx}	
\usepackage{amsmath}	
\usepackage{amssymb}	

\usepackage[FIGTOPCAP,Large]{subfigure}
\usepackage{cancel}

\title[WD disc evolution and environment]{Circularization of tidal debris around white dwarfs: implications for gas production and dust variability}

\author[Malamud Et Al.]{
	Uri Malamud,$^{1}$\thanks{E-mail: urimala@physics.technion.ac.il}
	Evgeni Grishin$^{2}$ and Marc Brouwers$^{3}$
	\\
	$^{1}$School of the Environment and Earth Sciences, Tel Aviv University, Ramat Aviv, 6997801 Tel Aviv, Israel\\
	$^{2}$Department of Physics, Technion - Israel Institute of Technology, Technion City, 3200003 Haifa, Israel\\
	$^{3}$Institute of Astronomy, University of Cambridge, Madingley Road, Cambridge CB3 0HA\\
}

\date{Accepted for publication in MNRAS, 14/12/2020.}

\pubyear{2020}

\begin{document}
	\label{firstpage}
	\pagerange{\pageref{firstpage}--\pageref{lastpage}}
	\maketitle
	
	\begin{abstract}
		White dwarf (WD) pollution is thought to arise from the tidal disruption of planetary bodies. The initial fragment stream is extremely eccentric, while observational evidence suggest that discs are circular or nearly so. Here we propose a novel mechanism to bridge this gap and show that the fragments can rapidly circularise through dust or gas drag when they interact with a pre-existing compact disc. We assume that the tidal stream mainly consists of small cohesive fragments in the size range 10-1000 m, capable of resisting the WD tidal forces, whereas the compact discs span a wide mass range. We provide an analytical model, accompanied by N-body simulations, and find a large parameter space in fragment sizes and orbital separation that leads to full circularization. Partial circularization is possible for compact discs that are several orders of magnitudes less massive. We show that dust-induced circularization inherently produces gas as tidal fragments collisionally vaporize the pre-existing dust along their path. We show that ongoing gas production has a higher probability to occur during the early stages of tidal disruption events, resulting from the fact that smaller fragments are the first to circularize. Intermittent gas production however becomes more likely as the tidal stream matures. This could explain why only a small subset of systems with dusty compact discs also have an observed gaseous component. Additionally, the interaction yields fragment erosion by collisional shattering, sputtering, sublimation and possibly ram-pressure. Material scattered by the collisions might form a thin dusty halo that evolves through PR drag, in compatibility with observed infrared variability.
	\end{abstract}
	
	\begin{keywords}
		white dwarfs -- planet–disc interactions
	\end{keywords}
	
	
	\section{Introduction}\label{S:Intro}		
	Between 25-50\% of white dwarf (WD) atmospheres are polluted with heavy elements \citep{ZuckermanEtAl-2003,ZuckermanEtAl-2010,KoesterEtAl-2014}, originating from accretion of planetary material \citep{DebesSigurdsson-2002,Jura-2003,KilicEtAl-2006,Jura-2008}. The details and especially the sequence of events leading to such an acceretion are however still not fully understood.
	
	Tens of WD systems are known to host debris, of either dust or gas (or both), which are in close vicinity to the WD, extending to $10^1-10^2$ WD radii \citep{Farihi-2016}. In terms of the WD tidal radius, observed debris are thought to be near or within the WD Roche limit, which is approximately equal to the progenitor star's main-sequence physical radius \citep{BearSoker-2015}. The fiducial Roche value is $R_{\odot}$, the Sun's present-day radius. The presence of orbiting dust is deduced from measurements of infrared excess \citep{RocchettoEtAl-2015}, while gas is inferred from metal emission lines or absorption features (see recent review by \citealp{ManserEtAl-2020}). The origin of material at such close proximity to the WD is clearly not primordial \citep{GrahamEtAl-1990}. During the post main-sequence evolution of the progenitor star it greatly expands and could swallow any pre-exiting planetary material up to a distance of a few AU \citep{MustillVillaver-2012}. Hence, the region that has been cleared prior to the WD stellar phase is many orders of magnitude larger than the domain of typically observed debris.
	
	The debris are instead thought to originate from planetary bodies which are perturbed by various potential mechanisms \citep{DebesSigurdsson-2002,BonsorEtAl-2011,DebesEtAl-2012,KratterPerets-2012,PeretsKratter-2012,ShapeeThompson-2013,MichaelyPerets-2014,VerasGansicke-2015,StoneEtAl-2015,HamersPortegiesZwart-2016,Veras-2016,PayneEtAl-2016,CaiazzoHeyl-2017,PayneEtAl-2017,PetrovichMunoz-2017,StephanEtAl-2017,SmallwoodEtAl-2018} to highly eccentric (and tidal-crossing) orbits, and are therefore subsequently tidally disrupted to form an initial eccentric circumstellar tidal stream of planetary fragments.
	
	During this initial formation phase, the morphology of the resulting tidal stream depends on the properties of the object being tidally disrupted. An asteroid would tidally disrupt to form a narrow ring of material, with very little spread in the orbital energy of its constituent fragments \citep{VerasEtAl-2014,MalamudPerets-2020a,NixonEtAl-2020}, unless it originates very far from the WD (as in Kuiper belt objects or Oort cloud objects) in which case the relative energy spread compared to the original orbital energy could be much greater. Larger objects that are the size of dwarf planets, moons or even terrestrial planets, would typically form a much more dispersed debris disc; i.e, the constituent fragments would have a large spread in orbital energy. The disc would be comprised of tightly bound interlaced elliptic eccentric annuli with semi-major axes extending from as little as $\sim$0.05 AU (eccentricities ranging from at least 0.9) to well beyond the original planet orbit \citep{MalamudPerets-2020a,MalamudPerets-2020b}. While these scenarios result in completely different debris discs, the outcomes nevertheless share a common morphological characteristic -- high eccentricity (eccentricities ranging from at least 0.9 and typically $\sim$ 1).
	
	Taking a leap forward in time, the studies of \cite{KenyonBromley-2017a,KenyonBromley-2017b} focus on the final formation sequence of the disc, when it is assumed that the disc attains a much more compact state (with maximum eccentricity of order $\sim 0.01$). Here, collisional grinding of large particles rapidly pulverize them to mere dust and (sublimated) gas. Such discs are circular or nearly so. They are relatively quiescent and are typically regarded as the canonical discs one often discusses in the context of observed WD discs with infrared excess \citep{Jura-2003}. In order to avoid confusion with evolving eccentric discs, throughout the paper we shall refer to this compact canonical disc configuration simply as 'compact discs'. 
	
	A recent study explores the possibility that highly eccentric, ring-like tidal debris, are also compatible with infrared excess evidence instead of compact discs \citep{NixonEtAl-2020}. Earlier work by \cite{DennihyEtAl-2016} also shows that it is possible to relax the typical assumption of a completely circular dusty compact disc, and that elliptical discs present an alternative interpretation. However, they acknowledge that the absence of ample long wavelength data makes it quantitatively difficult to draw important distinctions which discern between various high eccentricity models. Furthermore, both studies lack concrete quantitative modeling (while \cite{NixonEtAl-2020} provide a qualitative argument) accounting for the moderate eccentricites observed in a variety of discs (see Section \ref{SS:Discs} for details). We find it more likely that the extremely eccentric configurations proposed in \cite{MalamudPerets-2020b} (hereafter MP20) or \citep{NixonEtAl-2020} simply reflect an early or intermediate stage in the formation process of the compact discs that are normally associated with infrared excess.

	We are thus \emph{missing} an important link in between the initial formation stage and the compact disc stage. How do discs evolve from a state of such extreme initial eccentricity? \cite{VerasEtAl-2015} consider disc shrinkage through the drifting of small micron-to-cm sized particles by Poynting-Robertson (PR) drag, however it is unclear that this particle size range constitutes a significant mass fraction of the initial disc. Most studies actually advocate tidal fragments in the approximate size range of 10-1000m \citep{BrownEtAl-2017,KenyonBromley-2017a,Rafikov-2018,MalamudPerets-2020a}. For this size range, various arguments in \cite{MalamudPerets-2020a} and particularly in their Appendix A emphasize the potential importance of the Yarkovsky effect, which could be an important agent for circularizing the disc, however it remains to be properly explored in this context \citep{VerasEtAl-2015}. Alternatively, collisional cascade might potentially break the fragments on long evolutionary timescales, such that a more significant fraction of the disc could then evolve to circularize through PR drag \citep{WyattEtAl-2011}.
	
	In this manuscript we now suggest a third alternative, in which orbital dissipation and circularization of tidally formed fragments is facilitated through interaction with a pre-existing dusty or gaseous compact disc (or having both components). Our goal is to stipulate the conditions for this mechanism plays a significant role in the process of shrinking the disc (eccentricity) and in changing its fundamental properties. Few previous studies have began to explore the role of gas as a possible circularizing agent \citep{GrishinVeras-2019,OConnorLai-2020}, however in this paper we greatly extend the discussion and focus primarily on pre-existing dust, since it is observationally more prevalent than gas. It is estimated that between 1\% and 3\% of all single WDs with cooling ages less than around 0.5 Gyr possess circumstellar dust \citep{FarihiEtAl-2009,RebassaMansergasEtAl-2019}, while the occurrence rate for dust discs to have observable gaseous component (in emission) is only a few percent \citep{ManserEtAl-2020}. However, several newly-available pre-prints \citep{DennihyEtAl-2020,Gentile-FusilloEtAl-2020,MelisEtAl-2020} may have increased this fraction to about a third or more. Gaseous components are nevertheless rarer. The relative rarity of detectable gas may simply reflect an earlier stage in the circularization process. 
	
	In what follows, we will not only show that dust-assisted circularization is likely, but also that gas production is an inescapable by-product of the interaction between an eccentric tidal stream and a pre-existing dusty compact disc. In turn, we will show that detectable gas components are rarer since they correlate with (more probable to be observed in) the initial stages of disc circularization and evolution, when the tidal stream is still abundant with small fragments and gas production is continuous rather than intermittent. Finally we will show that if material is deposited far enough from the WD it can form an outer dust halo, compatible with observations of infrared variability seen in dusty WD discs.
	
	Recently, \cite{GrishinVeras-2019} (hereafter GV19) proposed an analytical formalism suitable for calculating gas-assisted circularization, based on an earlier work by \cite{GrishinEtAl-2019}. They suggest that small exo-Kuiper or exo-Oort like objects in the 0.1-10 km size range may be captured by a gaseous disc in the vicinity of a WD. Here we use a similar formalism but now investigate the evolution of realistic debris disc configurations, by using accurate initial conditions that emerge in the theoretical study of MP20, in addition to more realistic fragment sizes than in the GV19 study. More importantly, we present an extension of the GV19 formalism to include dust-assisted circularization in addition to gas-assisted circularization, as observational constraints dictate that the former should be the more central circularization agent. Our model assumptions and initial conditions are detailed in Section \ref{S:assumptions}. Our analytical model and additional numerical simulations are described in Section \ref{S:model}. The results are presented in Section \ref{S:results} and discussed in Section \ref{S:discussion}.
	
	Our model of course implies that a pre-existing compact disc is present, generated in the past by different mechanisms. The discussion elaborates on various formation possibilities, as well as the mass ratio between the tidal stream and the compact disc, which determines the outcomes of partial versus full circularization. A summary of the paper is provided in Section \ref{S:summary}.

	\section{Assumptions and initial conditions}\label{S:assumptions}
	We consider four vital elements of the model: (i) the mass and spatial domain of the dust/gas components in a pre-existing compact disc ; (ii) the size of interacting eccentric fragments; (iii) the orbit of eccentric fragments; and (iv) the point where fragments intersect the compact disc. They are respectively discussed in the following subsections.
	
	\subsection{Disc masses and spatial extent}\label{SS:Discs}
	Dust discs are commonly characterized by a typical infrared excess that can be explained via circumstellar dust at a temperature of around 1000 K. In the standard technique for modelling infrared excess \citep{Jura-2003} it is assumed that the WD illuminates a passive, opaque, flat circumstellar annulus where the incident optical energy is re-radiated in the infrared. This yields a typical declining disc temperature as a function of radial distance. In turn this constrains the flux, and when compared with the observations, the inner and outer radii of the annulus can be extracted. There is a considerable degeneracy between the disc inclination and the disc radial extent. For example in \cite{DennihyEtAl-2016} the disc width of WD EC 05365 can be modelled with varying inner and outer radii in the range of approximately $\sim 0.4-0.9 R_\odot$ (see their table 3). In some cases the inclination can be constrained. E.g. for a transiting system as in WD 1145+017 \citep{VanderburgEtAl-2015}, the inclination must be high (most observational studies define $i=90^{\circ}$ for when the disc is roughly edge-on to account for the transit, whereas we consider $i=0$ in this case) and \cite{ZhouEtAl-2016} constrain the dust position from infrared excess to $\sim 0.9 R_\odot$. The inclination may also be constrained in discs that have both dust and emitting gas \citep{GansickeEtAl-2006,ManserEtAl-2016b}. For a given inclination of 70$^{\circ}$, \cite{BrinkworthEtAl-2009} constrain the dust disc of SDSS J1228+1040 to range between 0.2-1.2 $R_\odot$.
	
	While up to half of all WDs have atmospheres polluted with heavy elements \citep{KoesterEtAl-2014}, only a small fraction of order few \% are observed to possess circumstellar dust \citep{FarihiEtAl-2009,RebassaMansergasEtAl-2019}. Unless most WDs accrete material directly, which is unlikely \citep{DebesSigurdsson-2002}, accretion has to be mediated by a disc, suggesting that most discs simply evade detection. Indeed \cite{RocchettoEtAl-2015} study the distribution of infrared fractional luminosities as a function of the WD cooling age and show that they are significantly less than the maximum value allowed for flat discs. In other words, these discs seem to be less extended compared with their allowed theoretical range, and for WDs with cooling age under $\sim$ 0.5 Gyr there is no increase in the fractional luminosities as the WDs mature, despite the expected inward migration of the inner sublimation zone. Evading infrared detection therefore must involve insufficient surface area of the emitting dust grains. \cite{RocchettoEtAl-2015} suggest three potential compact disc configurations that could escape detection: (1) high fraction of gas / completely gaseous discs \citep{Jura-2008} ; (2) narrow (in radial extent) opaque dust rings \citep{FarihiEtAl-2010}; (3) optically thin dust discs \citep{BochkarevRafikov-2011}. Similar ideas were proposed by \cite{BonsorEtAl-2017}. We will return to these points later on in Section \ref{SS:ComparableMasses}.
	
	This last paragraph importantly indicates that the classical view of a pre-existing dust disc as a Saturn-like, geometrically thin and optically thick disc is far from certainty. Rather, at least in WDs with cooling age less than 0.5 Gyr, the majority of discs \emph{do not} conform with the canonical compact disc configuration, as evident by the small fraction of IR excess detections, despite the favourable observational bias in this class of WDs. Nevertheless, in the context of our paper we will assume that pre-existing dust discs initially adhere to the classical view, but we will also discuss how their structure and thus detectability can change in the aftermath of a tidal disruption event. Our assumed inner and outer disc radii for pre-existing dust are 0.2 and 1 $R_\odot$ respectively, in accordance with the inferred observational constraints previously discussed. 
	
	The first clue for uncovering disc masses (both gas and dust) comes from the average WD accretion rates. The latter can be estimated via the WD observed photospheric metal abundances. Using theoretical estimates for the size of the outer convective zone of the WD, the amount of polluting mass may be derived from the aforementioned metal abundances, and divided by the theoretical estimates for the heavy elements sinking times to yield the average accretion rate (see \cite{Koester-2009} and references therein for both the former and the latter). The ensuing average accretion rates fall within a relatively wide range of $10^5-10^{11}$ g $ \times $ s$^{-1}$ \citep{FarihiEtAl-2009,FarihiEtAl-2010,KoesterEtAl-2014}. As a heuristic approach, let us assume a constant accretion rate throughout the lifetime of the disc, which is empirically estimated by \cite{GirvenEtAl-2012} as $\sim 10^4-10^6$ yr. The total disc mass $M$ then falls within an even wider range of $10^{16}-10^{24}$ g. The upper limit estimate of $10^{24}$ g is agreeable with the maximal mass observed in He-dominated WDs \citep{FarihiEtAl-2010,XuJura-2012,GirvenEtAl-2012}. We consider this mass range appropriate for dust discs. In the following paragraphs we will also discuss gas, its radial extent and possible mass range.
	
	Gas discs are typically found in combination with observed infrared excess, i.e in dust discs that have a gaseous component. They are extremely rare with only 12 known WD discs that contain a gaseous component \citep{ManserEtAl-2020}, hence pre-existing gas discs are not central to the ideas presented in this paper, but we nevertheless thoroughly discuss them for completion and since they help us constrain the spatial extent of compact discs. In 7 of the 12 aforementioned discs the presence of gas has been detected via the Doppler-broadened, double-peaked line emission of the Ca II 8600 \r{A} triplet, which results from the Keplerian rotation of a flat disc that is photo-ionised by the WD. In the other five discs the gas was detected via absorption features. For the moment, the underlying differences between gas emission and absorption features in these systems is not well understood \citep{ManserEtAl-2020}. Only one WD (SDSS 1228+1040) stands out in having both emission and absorption \citep{GansickeEtAl-2012}. To contrast, WD 1145+017 has a similarly highly inclined (i.e edge-on) disc and has been heavily monitored (see \cite{CauleyEtAl-2018,RappaportEtAl-2018,IzquierdoEtAl-2018,KarjalainenEtAl-2019,XuEtAl-2019,FortinArchambaultEtAl-2020} and numerous older references therein), yet it does not appear to have emitting gas. The situation is similar with SDSS J1043+0855 \citep{ManserEtAl-2016b,MelisDufour-2017}. The low statistics involved currently make it difficult to speculate on the root cause that differentiates between these systems. We also note that in the time interval between the submission and revision of this manuscript, several pre-prints \citep{DennihyEtAl-2020,Gentile-FusilloEtAl-2020,MelisEtAl-2020} reported (in total) 15 new WD discs with gaseous emission, tripling the number of previously known discs with gas. We mention these studies for completion. 
	
	Some spatial constraints on the shape and radial extent of the gas may be obtained by modelling the gas emission. Significant asymmetry in the line profiles suggests that the gas component has non-negligible eccentricity, ranging between $\sim 0.2-0.4$ \citep{GansickeEtAl-2006,GansickeEtAl-2008,MelisEtAl-2010}. If modelled via a series of co-aligned elliptical orbits of identical eccentricity, as aforementioned, the radial distribution of the gas may be constrained. Correcting the velocities by accounting for the dust disc best-fit inclination angles, previous estimates have placed the gas at a distance of around $\sim$0.15-1.2 $R_\odot$ \citep{GansickeEtAl-2006,GansickeEtAl-2008,MelisEtAl-2010}.
	
	Constraints on the spatial extent of the circumstellar gas from absorption features have so far only been modelled for one system, that of WD 1145+017 \citep{FortinArchambaultEtAl-2020}. Despite several approximations used in the construction of this model, it places the gas at a compatible distance of $\sim$0.2-0.5 $R_\odot$.
	
	Finally, over long periods, looking at the time evolution of the gas emission and line morphology shows smooth progression from a red-dominated asymmetry to a blue-dominated one. The morphology of the peaks also changes, with the blue-shifted peak becoming stronger and sharper, while the redshifted peak becomes shallower and weaker and extends to higher velocities. Overall this evolution is commensurate with relativistic apsidal precession of an eccentric gas component in the gravitational potential of the WD \citep{ManserEtAl-2016a,CauleyEtAl-2018,FortinArchambaultEtAl-2020}. This places interesting constraints for the precession periods, ranging from 4.6 yr for WD 1145+017 \citep{CauleyEtAl-2018,FortinArchambaultEtAl-2020} to 27 yr for SDSS J1228+1040 \citep{ManserEtAl-2016a}. In the latter study, \cite{ManserEtAl-2016a} find that precession periods of 1.54, 27.8 and 134 yr correspond to orbits with semi-major axes of 0.2, 0.64 and 1.2 $R_\odot$, respectively (adopting a WD mass of 0.7 $M_\odot$, and in the limit of small eccentricity). These estimates are in approximate agreement with previously mentioned gas regions inferred through other techniques.
	
	Taken at face value, the aforementioned constraints imply that when gas is observed, it inhabits the inner portions of WD compact discs, and may extend approximately up to $R_\odot$. As already mentioned, this line of evidence is also very compatible with the radial extent evident in dust discs. In both cases $R_\odot$ is suggested as an approximate upper limit, in excellent agreement with the typical WD Roche limit. We will take the most conservative approach, allowing the gas, if it exists, to extend only to $\sim 1R_\odot$, which is similar to the approach in the study of \cite{vanLieshoutEtAl-2018}. For the gas inner edge we choose 0.2 $R_\odot$, based on the aforementioned observational constraints.

	
	
	The mass of the gas in the debris is even more uncertain from observations. The mass ratio of gas-to-dust in the disc is essentially unconstrained. According to \cite{Veras-2016} it ranges from $10^{-5}$ to unity. Given the aforementioned dust disc mass range of $10^{16}-10^{24}$ g, and adopting $10^{-5}$ as the lowest possible (most limiting) gas-to-dust mass ratio, the gass mass ranges from $10^{11}-10^{24}$ g. We nevertheless note in the interest of accuracy, that purely gaseous discs of masses $10^{23}-10^{24}$, in the upper range suggested above, are not likely to apply to polluted WDs. This mass is rather typical of cataclysmic variable discs, and requires an input mass accretion rate of at least $10^{13}$ g$\times$s$^{-1}$ (Figure 3 in \cite{Hameury-2020}), which is both incompatible with WD disc accretion rates, as noted previously, and should also appear as a bright X-ray source, contrasting with observations.
	
	In combination, what we know about both dust and gas discs seems to indicate that material is roughly delineated between 0.2-1 $R_\odot$, which is what we will assume in our baseline model. In our forthcoming analysis we will further assume for simplicity that the material in the disc initially spreads throughout the entire radial extent of the disc rather than confined to a narrow region. We treat the mass range as a free parameter in the range suggested for dust mass, i.e covering 8 orders of magnitude between $10^{16}-10^{24}$ g. Masses under $10^{16}$ g are ignored as they are too small to yield effective circularization as we shall see in Section \ref{S:results}. Masses close to $10^{24}$ are probably rare or inapplicable, yet we consider them for completion.
	
	\subsection{Fragment sizes}\label{SS:FragSizes}
	The GV19 study considers gas interactions with objects in the 0.1-10 km size range, appropriate to their Oort-cloud origin. We however consider tidally disrupted fragments, and therefore regardless of the size of their progenitor, the correct size range should ensue from the intrinsic properties of the progenitor as well as its precise origin and how deeply it penetrates the tidal sphere. Overall we suspect that the fragment size range is reduced compared to the GV19 study. We refer to the \cite{BrownEtAl-2017} study, in which tidal fragments emerging from tidal disruptions of larger parent objects and are considered as monolithic objects with some internal strength, when their size is estimated at tens to hundreds of m. Similar estimates of fragment sizes, in the range 10-1000 m, are given in the \cite{KenyonBromley-2017a} study (see e.g. their Figure 1).
	
	In a different study, \cite{Rafikov-2018} considers the vertical tidal collapse of a large planetesimal as it deeply penetrates the tidal sphere of a WD. In this context, he envisions the planetesimal as a collection of fragments which collide with each other as they collapse inwards towards the orbital mid-plane. These collisions start off as catastrophic, reducing fragments sizes in a collisional cascade. However, below a minimum size of roughly $\sim$100-300 m, depending on the exact material \citep{LeinhardtStewart-2009}, the collision velocity for catastrophic fragmentation rather increases as material strength becomes more important. \cite{Rafikov-2018} thus hypothesizes that the final size distribution of fragments resulting in this process should peak around the size at which the collision velocity leading to catastrophic disruption is minimized. This size is estimated to be 0.1-1 km in accordance with the \cite{LeinhardtStewart-2009} findings.
	
	The study by MP20 also performed detailed tidal disruption simulations, finding that as much as 70-80\% of the tidal fragments have relatively short rotation periods less than twice the known 2.2 h cohesion-less asteroid spin-barrier (only sufficiently small and cohesive/monolithic asteroids are able to spin faster than this limit). Hence, the cohesion-less asteroid spin-barrier might also serve as clue to the potential size of the tidal fragments. Asteroids spinning faster than the spin-barrier fall in the size range of 150-300 m according to \cite{PravecEtAl-2002}. Given all these independent estimates, we treat the fragment sizes as a free parameter ranging some 2 orders of magnitude, between 10-1000 m. Smaller or larger fragments are ignored since we consider their initial mass fraction in the stream to be negligible.
	
	
	Throughout this paper we consider only fully formed tidal stream. The MP20 and \cite{VerasEtAl-2014} studies show that several orbits (of the progenitor) are required for the stream to fully evolve. Before this, the fragments could be more clumped up and not spread uniformly throughout the stream, and if the pericentre is far from the WD, partial disruptions result in large fragments that require many flybys before they finish dissecting to their smallest possible constituents. However since the stream formation timescale is relatively short, we ignore this phase entirely.
	
	\subsection{Initial fragment orbits}\label{SS:InitialFragDistances}
	Simple analytical arguments given in Section 1 of \cite{MalamudPerets-2020a} provide compelling evidence that the tidal fragments following a disruption event should occupy diverse initial semi-major axes, based on the precise origin and size of the progenitor object. For example, a small km-sized asteroid originating from a minimal distance of at least several AU (required in order to survive the host star's post main-sequence evolution, see \cite{MustillVillaver-2012} and \cite{MustillEtAl-2014}) would disrupt to form a ring of debris on the original asteroid semi-major axis. Such a ring would have a very tiny spread in the orbital energies of the fragments \citep{VerasEtAl-2014,MalamudPerets-2020a,NixonEtAl-2020}.
	
	A common misunderstanding is to assume that the same outcome applies to all small asteorids. For example, consider an analogue Kuiper belt object akin to recently visited Arrokoth \citep{GrishinEtAl-2020a,McKinnonEtAl-2020}. It orbits its main-sequence host star at a distance of several tens of AU, and orbital expansion primarily during the AGB phase of the host star would likely drive its semi-major axis to $\sim$150 AU when the star reaches the WD phase (depending on the host star's mass, see e.g. \cite{MalamudPerets-2017a} and \citep{MalamudPerets-2017b}). Since its size is of the order $\sim 10^1$ km, it is large enough and distant enough to form a moderately dispersed debris disc rather than a ring, when it tidally disrupts upon reaching a distance of $R_{\odot}$ from the WD (its outer tidal sphere limit). It may even marginally produce unbound debris (as evident in figure 2 of \cite{MalamudPerets-2020a}), while the innermost semi-major axis $a$ in the tidal stream (see their equation 3) would be halved. If its close approach distance is smaller than $R_{\odot}$, then the fragment dispersion increases considerably. Thus, a non ring-like outcome is clearly plausible and even likely in small asteroids that originate in exo-Kuiper belts or beyond.
	
	As we consider tidal disruptions of larger objects, in the size range $10^2-10^3$ km or more (or else the pericentre distance is much less than $R_{\odot}$), the tidal fragments transition from forming a ring configuration, or a disc with moderate orbital dispersion as previously shown, to a bi-modal semi-major axis distribution \citep{MalamudPerets-2020a}. In the latter configuration, half of the progenitor material becomes unbound and the other half becomes tightly bound to the WD, with $a$ as little as 0.05 AU. The fragment semi-major axis distribution is approximately that given by Figure 2 of MP20, the peak in the distribution ranging between 0.1 to 1.5 AU for progenitors of mass 1$M_{\oplus}$ to 10$^{-4} M_{\oplus}$ respectively.
	
	We conclude that different tidal disruption scenarios lead to very different semi-major axes of the tidal fragments interacting with the pre-existing compact disc. To quantify the range of $a$, let us recap the aforementioned cases as discussed above, assuming the pericentre $q$ at $R_{\odot}$ for our analysis: (i) disruption of large progenitors result in a bi-modal fragment distribution with $a$ between $10^{-1}-10^0$ AU; (ii) asteroids originating at several AU form rings. Hence fragments would orbit on the original progenitor semi-major axis of few $10^0$ AU; (iii) typical Kuiper belt analogues form moderately dispersed debris if their size is $\sim$10 km. If smaller, then their debris remain close to the original progenitor orbit. Overall the likely fragment semi-major axis distribution is of order few $\sim 10^1-10^2$ AU; (iv) typical Oort cloud analogues with original orbits in the range $10^3-10^5$ AU form anything from moderately dispersed debris ($\sim$1 km objects at $10^3$ AU) to fully bi-modal debris ($\sim$10 km at $10^3$ AU, or $\sim$1 km objects at $10^4$ AU and beyond). In a bi-modal disruption regime the fragment semi-major axes converge onto a value of $q^2/2L$ \citep{MalamudPerets-2020a}, where $L$ is the progenitor radius. Hence we expect in all of these cases that most of the fragments will have $a$ around $\sim 10^3$ AU (while very few fragments may indeed extend to the original orbit or more, however their numbers would be entirely negligible). 
	
	We note that the cases discussed in the previous paragraph were analysed assuming a pericentre $q$ of $R_{\odot}$. If a lower $q$ is used in the same analysis the fragment semi-major axes are further reduced since deeper disruptions are necessarily more dispersive. In conclusion, we treat the semi-major axis of bound fragments as a free parameter, ranging between $10^{-1}-{10^3}$ AU. Tidal fragments would rarely have $a$ larger than about $10^3$ AU hence this range is strongly constrained.
	
	\subsection{Intersection between tidal fragments and compact disc}\label{SS:FragDicIntersection}
	The intersection point or points between the tidal fragments and the pre-existing compact disc will occur at roughly the pericentre of the tidal stream (i.e. the pericentre of the progenitor object). Our approach for selecting the preicentre distance is similar to that of the MP20 study. They focus primarily on tidal disruptions in which the progenitor's pericentre is Roche-grazing and corresponds to 1$R_{\odot}$, an approximate outer-limit. However, their complete investigation includes three choices for the pericentre which cover the full range of distances relative to $R_{\odot}$: deep, intermediate and grazing.
	
	Here we also consider three cases for the pericentre distance: an inner intersection of 0.2$R_{\odot}$ (i.e intersecting the compact disc along its inner-edge), an intermediate intersection point of 0.6$R_{\odot}$ and an outer, grazing intersection of 1$R_{\odot}$. We make the simplification that the argument of pericentre is zero, so that the fragments have one intersection point with the compact disc rather than. Later in Section \ref{SS:analytic} we show that if it were not the case, our results would only change quantitatively within a factor unity, which is why such an assumption is judicious. We refer the reader to Figure \ref{fig:4}, where a schematic diagram shows the intersection point between a crossing fragment with the compact disc. Most of the details in this diagram, however, will only be elaborated on in Section \ref{S:discussion}.

	\section{Model}\label{S:model}
	\subsection{Analytical formalism}\label{SS:analytic}
	We consider a generic profile for a pre-existing compact disc, which is typically composed of dust, and in some cases could have a gaseous component as well. We then describe the circularization mechanism due to the drag forces between fragments in the tidal stream and the pre-existing compact disc, with the goal of calculating the circularization timescale. The model was initially developed by GV19 strictly for gas discs, however here we extend the model discussing how dust-induced drag may be equivalent, and therefore applicable in the limit of very high velocities.
	
	Regardless of disc material type, we assume in accordance with the observational constraints Section \ref{SS:Discs} that the radial extent of the compact disc ranges from $r_{\rm in}=0.2R_\odot$ to $r_{{\rm out}}=R_\odot$. Our surface density profile is given by \begin{equation}
	\Sigma(r)=\Sigma_{0}\left(\frac{r}{R_{\odot}}\right)^{-\beta},
	\label{eq:sigma}
	\end{equation}
	
	\noindent where $\beta$ is an arbitrary exponent. For both dust and gas we shall take $\beta=3/2$, as follows. When considering gaseous WD discs, \cite{MetzgerEtAl-2012} find that the gaseous surface density $\Sigma_{g}(r) \propto r^{-n-1/2}$, where $n$ describes the viscosity power law $\nu (r) \propto r^n$. For an optically thin gas disc with $T(r) \propto r^{1/2}$, they take $n=1$ and $\beta=3/2$. When considering dusty WD discs, the radial profile is essentially unknown. However, in young proto-planetary discs the canonical minimal mass Solar nebula predicts a power law of $\beta=3/2$ \citep{wei77}. This scaling is physically justified if the solids distribute the specific angular momentum $\ell$ in the disc as $\ell \propto \sqrt{a}$, similarly to the Keplerian motion of the planets. Without any other knowledge of the structure of compact discs around the WD, we also follow this prescription. Thus, $\beta=3/2$ is an acceptable choice for both material types. Additionally, we also note that the model is not very sensitive to different choices of $\beta$ in the range 1-2.
	
	The total mass of the disc is then
	\begin{align}
	M_{}  \approx \intop_{r_{\rm in}}^{r_{\rm out}}\Sigma_(r)2\pi rdr\approx \frac{4\pi}{3}\Sigma_0R_\odot^{2},
	\label{Mdisc1}
	\end{align}
	
	\noindent with the surface density constant
	
	\begin{align}
	\Sigma_0\approx 5\cdot 10^{-4} \rm{g\ cm^{-2}}\left(\frac{M}{10^{19}{\rm g}}\right). \label{eq:sigma0}
	\end{align}
	
	Consider a tidal fragment of size $R$, density $\rho$ and mass $m$, orbiting a WD of mass $M_{\rm WD}$ with a semi-major axis $a$ and a grazing pericentre $q=a(1-e)\ll a$. \cite{GrishinEtAl-2019} calculate the energy loss due to drag during a single passage (see justification in Appendix \ref{A:IntersectionPoints}) through the disc at pericentre if the coincident material in the disc is gaseous. They further assume, in order to minimize the number of free parameters, that the fragment crosses the disc face-on, neglecting the inclination $i$ which contributes an order unity pre-factor. This means that our forthcoming analytical model circularization timescale may be considered as a lower limit result. The drag induced is simply the familiar aerodynamic gas drag. The energy loss is therefore
	
	\begin{equation}
	\Delta E=-\frac{\pi}{2}C R^2 \Sigma(q) v_{\rm rel}^2 \label{eq:de},
	\end{equation}
	
	\noindent where $C$ is the drag coefficient, $v_{\rm rel}^2$ is the relative velocity between the fragment and the compact disc and $\Sigma(q)$ is replaced with $\Sigma_{\rm g}(q)$, the gas surface density. For a highly eccentric orbit, the velocity is approximated by the escape velocity $v_{\rm rel}^2 = 2GM_{\rm WD}/q$, where $M_{\rm WD}$ is the mass of the WD.
	
	The drag coefficient measures the strength of the pressure difference in a gaseous medium. For low velocities (or low Reynolds numbers) the drag force is in the linear (Stokes) regime, which means that $C \propto v_{\rm rel}^{-1}$. Conversely, passage at extremely high velocity is very fast compared to the response of the gas, and the drag force is quadratic and proportional to $v_{\rm rel}^2$ (ram-pressure regime), which means that $C$ is of order unity.
	
	What happens when the coincident material is dusty rather than gaseous? Here the analogy of a Stokes drag law is unclear and it is not certain that dust drag is even active for low velocities. However, for very high velocities the details of the ambient medium are unimportant and the drag law behaves as in the ram-pressure regime with $C \approx 1$. Indeed, analogies between collisionless and gaseous medium have been proposed in the context of dynamical friction, or gravitational drag, either in 3D geometry \citep{RaphaeliEtAl-1980,Ostriker-1999}, or in 2D slab geometry \citep{MutoEtAl-2011, GrishinPerets2015}. While significant differences were present for the subsonic case, the hypersonic case of very high velocity essentially behaves universally, regardless of the property of the medium. We thus conclude that in the limit of high velocities, Equation \ref{eq:de} is (to order unity) indifferent to the type of compact disc material coincident with the passing fragments. Summarizing the previous paragraphs, both $\beta$ and $C$ are approximately independent of the coincident material type. From this point onwards, this section will extend the GV19 study while treating both dust and gas via a generic, identical formalism.		
	
	In order to complete their circularization, the tidal fragments must interact with a sufficient amount of material to move from their eccentric orbital energy $E = -GM_{\rm WD}m/2a$ to a contracted and circular energy state $E_{\rm circ} = -GM_{\rm WD}m/2q$, where $m$ is the fragment mass.
	
	The fractional energy change per orbit is given by
	\begin{equation}
	\mathcal{Q}^{-1}\equiv \left|\frac{\Delta E}{E}\right|=\frac{3C\Sigma(q)}{2\rho R}\frac{a}{q}, \label{eq:Q}
	\end{equation}
	
	\noindent where the quantity $\mathcal{Q}$ is the "quality factor" that measures how many (equivalent to the original) orbits the object will do before it will lose a substantial amount of energy and will be circularized. The definition is analogous to the tidal factor $\mathcal{Q}$ that is frequently used to describe tidal dissipation in two-body systems \citep{GoldreichSoter-1966, Hut-1981}, only that here the source of the dissipation is the induced drag. Note that $E$ is given by the initial orbital energy while in reality $a$ changes as a function of time. $\mathcal{Q}$ is therefore only a zeroth-order approximate to the number of orbits. In Section \ref{S:results} we however compare this simple expression with numerical computations and show that it is indeed accurate to order of magnitude.
	
	For full circularization within one to a few orbits, or $\mathcal{Q} \sim 1$, the fragment should be sufficiently small, i.e. $R \lesssim 1.5 C (a/q) \Sigma/\rho$. Larger fragments with $\mathcal{Q}\gg 1$ will take longer to fully circularize. The typical circulariztion time is $\tau_{\rm circ}=\langle |E/\dot{E}| \rangle$, where the brackets indicate averaging over the orbit. Since the loss of energy is impulsive and occurs at the pericentre, the energy loss can be estimated as $\dot{E} = \Delta E / \Delta t$, where $\Delta E$ is given in Eq. \ref{eq:de}, and $\Delta t=P=2\pi \sqrt{a^3/GM_{\rm WD}}$ is the orbital period, which leads to the circularization time simply being
	
	\begin{equation}
	\tau_{\rm circ} =  \mathcal{Q}P = \frac{4\pi \rho R q  }{3C \Sigma}\sqrt{\frac{a}{GM_{\rm WD}}}.  \label{eq:tcirc}
	\end{equation}
	
	Note that the timescale increases with the separation as $a^{1/2}$, meaning that distant fragments will require more time to circularize, however it will require less iterations since $\mathcal{Q}\propto a^{-1}$.
	
	\subsection{Caveats and restrictions}\label{SS:restrictions}
	There are three restrictions entailed in our simple model. The first restriction relates to time. Drag-induced dissipation is active as long as the ratio between the orbital period and the disc lifetime is small. The most distant tidal fragments, with $a$ around $10^3$ AU (Section \ref{SS:InitialFragDistances}) have orbital periods of the order of several $10^4$ yr, which roughly correspond to the lower limit empirical disc lifetime of $10^4$ yr \citep{GirvenEtAl-2012}. Hence, if the pre-existing compact disc lifetime is not affected by the interaction itself (a point which is further discussed in Section \ref{SS:ComparableMasses}), we may conclude that most fragments are capable of significant drag-induced circularization, however they require $\mathcal{Q}\lesssim 1$. Luckily $\mathcal{Q}$ is inversely proportional to $a$ as we have shown, and indeed yields low values. Conversely, tightly orbiting fragments with $a$ of the order of 1 AU have an orbital period of merely 1 yr. In line with the previous argument, they can have $\mathcal{Q}$ as high as $10^4$ and still circularize.
	
	The second restriction relates to conservation of energy. When we say that the drag coefficient $C$ is of order unity, it is the equivalent of saying that the pre-existing dust or gas attains the same velocity as the fragment. In other words, the energy loss in the fragment's orbit (Equation \ref{eq:de}) must be compensated by energy gain of the interacting compact disc material. One other possibility is that some of the energy goes to collisional vaporization of pre-existing dust, however it is easy to show that in the limit of the very high relative velocities in our scenario, $E_{vap} \sim 1.5*10^{11}$ erg $\times$ g$^{-1}$ (for SiO2, see \cite{Podolak1988}) is not very energy-expensive which means that energy conservation implies gaining roughly identical velocity (although not necessarily in the same direction). This in turn implies that the passage of a single fragment both scatters material and carves a gap in the compact disc. The gap has to be filled prior to interacting with another fragment, otherwise the above model is incorrect. The proper filling time is quantitatively discussed in Section \ref{SS:GapFilling}.
	
	The third restriction relates to conservation of momentum. Since the collisional interactions conserve angular momentum, keeping the compact disc angular momentum pointing in the same direction cannot work unless $M \gg M_{\rm progenitor}$, or else compact disc and progenitor object are co-orbital from the beginning. Even more importantly, if instead $M \lessapprox M_{\rm progenitor}$, this would entail a catastrophic fate to the part of the compact disc which interacts with the tidal stream, and diffusion would not be able to continue filling the gaps as previously discussed. Therefore, full circularization is necessarily truncated and only partial circularization can occur. We nevertheless note that even in the case of partial circularization our model is still very useful, as we discuss in the next section.
	
	Finally, we note that our model does not directly consider any effects from shocks, although the compact disc may indeed have a gaseous component, and the interaction is obviously extremely hypersonic. This aspect is beyond the scope of this paper, having already introduced many novel aspects, however it should be explored further in future dedicated studies. 
	
	\subsection{Partial circularization}\label{SS:PartialCircularization}
	The calculation in Section \ref{SS:analytic} assumes that the mass of the compact disc far exceeds that of the tidal stream, such that circularization can proceed fully. However, since the compact disc forms from bodies that are likely ejected to tidal-crossing orbits from the same reservoir of planetesimals as the tidal stream itself, it is probable that the fragments often encounter a disc whose mass is insufficient for complete circularization. In this case, the limit in dust mass to collide with leaves the fragments on shrunken orbits ($a_\mathrm{new}$) that lie within their initial semi-major axis $a$ and the pericentre $q$. Let us assume for simplicity that the fragments all have the same size and mass, and therefore we may equate the total orbital energy of the stream plus the energy loss from Equation \ref{eq:de} to the new total energy of the stream:
	
	\begin{equation}\label{eq:EnergyPartial}
		\frac{-GM_{\rm WD}M_{\rm progenitor}}{2a}-\frac{\pi}{2}C v_{\rm rel}^2 \Delta M = -\frac{-GM_{\rm WD}M_{\rm progenitor}}{2a_{\rm new}},
	\end{equation}
	
	\noindent where $\Delta M$ is some fraction of the compact disc mass $M$ that the stream has collided with, $M_{\rm WD}$ is the WD mass and $M_\mathrm{progenitor}$ is the mass of the tidal stream progenitor. Equation \ref{eq:EnergyPartial} then simplifies to
	
	\begin{equation}\label{eq:partial}
	\frac{a}{a_{\rm new}} = 1+2\pi C \frac{a}{q} \frac{\Delta M}{M_{\rm progenitor}}, 
	\end{equation}
	
	 Equation \ref{eq:partial} is valid up to the last stages of circularization, when the differential velocity shifts away from $v_{\rm rel}^2 = 2GM_{\rm WD}/q$ to a lower value (as further discussed in Section \ref{S:results}). Interestingly, it indicates that while complete circularization of the stream requires a mass that is at least comparable to that of the progenitor, significant partial circularization already occurs for very small disc masses. As an example, shrinking the initial orbit from $a$ to $a/2$ only requires a disc-to-stream mass ratio of $\Delta M / M_\mathrm{progenitor} = q / (2\pi C a)$. This amounts to tiny mass fractions of merely $10^{-2} - 10^{-7}$ for the stream semi-major axis being 0.1 to 1000 AU respectively (and $q$ in the range $0.2-1~R_\odot$). Hence, while complete circularization is only expected to occur in a limited fraction of cases, nearly all tidal streams that encounter a pre-existing compact disc will experience significant drag-induced orbital shrinkage.
	
	The likely prevalence of partial circularization begs the question of what happens after this process completes. We note that since the tidal stream is composed of different-sized fragments, partial circularization of the fragments will result in orbital differences based on fragment size. This in turn will lead to different apsidal precession rates and can ultimately facilitate catastrophic collisions among remaining fragments in the stream until constituent particles are sufficiently small to be transported by PR drag. We do not work out the detailed consequences of partial circularization here but will dedicate a future work to this process.
	
	Finally, full circularization occurs when we set $a_{\rm new}=q$ in Equation \ref{eq:partial}, which works out to give
	
	\begin{equation}\label{eq:FullCircularization}
		\Delta M = \frac{1}{2 \pi C} \left( M_{\rm progenitor}-\frac{q}{a} \right) \approx \frac{M_{\rm progenitor}}{2 \pi C}. 
	\end{equation}
	
	Equation \ref{eq:FullCircularization} is tantamount to the simple statement that the fragments in the tidal stream should encounter their own mass (within a factor of unity) inside the compact disc for them to achieve full circularization.	
	
	
	
	\subsection{Numerical simulations}\label{SS:numeric}
	In order to test and better quantify the analytical results, we employ numerical N-body simulations of debris on initially eccentric orbits passing through a pre-existing disc. For the N-body integration, we use the publicly available code \texttt{REBOUND} \citep{ReboundMain}. We use \texttt{IAS15}, a fast, adaptive, high-order integrator for gravitational dynamics, accurate to machine precision over a billion orbits \citep{ReboundIAS15}. The drag force is modelled with a constant drag coefficient $C$ equals unity, as discussed in detail in Section \ref{SS:analytic}.
	
	The surface density is prescribed according to Equations \ref{eq:sigma} and \ref{eq:sigma0}. For the mid-plane volume density, it is given by $\rho(z=0,r)=\Sigma(r)/h/\sqrt{2\pi}$, where $h$ is the scale height of the disc and is set to the fiducial value of one percent of the radial distance, $10^{-2} r$. The local density in the $z$ plane is set by a Gaussian profile, $\rho(z,r)=\rho(z=0,r) \exp (- z^2/2h^2)$. While this treatment is typical of gas discs (taking after the GV19 study), we apply it here in a broader sense to both dust and gas.
	
	The disc is assumed to rotate with a Keplerian velocity of a circular orbit and is assumed to far exceed the mass of the fragment stream, such that its circularization can fully proceed as discussed in Section \ref{SS:restrictions}. Numerical results would be presented and compared to analytical ones in the following section, where we explore the full parameter space. 
	
	
	\section{Results}\label{S:results}
	The analytical model described in Section \ref{S:model} allows us to estimate the typical time for circularization as $\tau_{\rm circ}$. Here we explore the dependence of $\mathcal{Q}$ and $\tau_{\rm circ}$ on the size and orbit of the fragments, in addition to the disc mass $M$. We consider the canonical fragment density of $\rho= 3 \ \rm g \ cm^{-3}$ and WD mass of $M_{\rm WD} = 0.6~M_\odot$.
	
	\begin{figure}
		\subfigure[$q=1 R_\odot$] {\label{fig:1a}\includegraphics[width=\columnwidth]{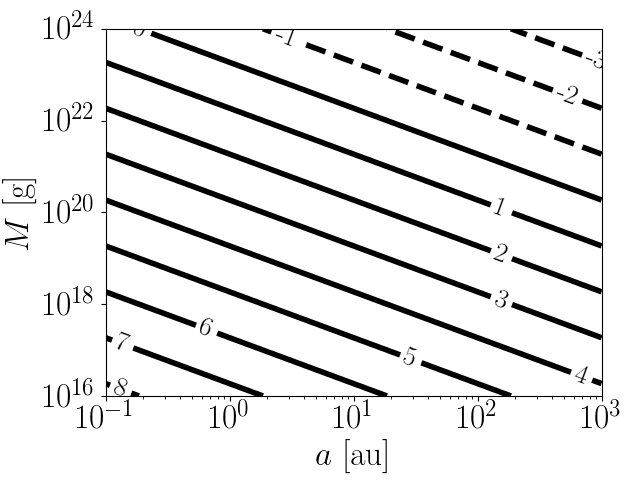}}
		\subfigure[$q=0.6 R_\odot$] {\label{fig:1b}\includegraphics[width=\columnwidth]{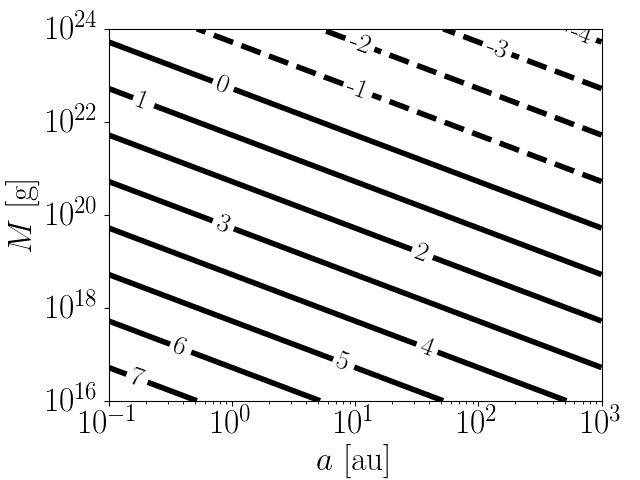}}
		\subfigure[$q=0.2 R_\odot$] {\label{fig:1c}\includegraphics[width=\columnwidth]{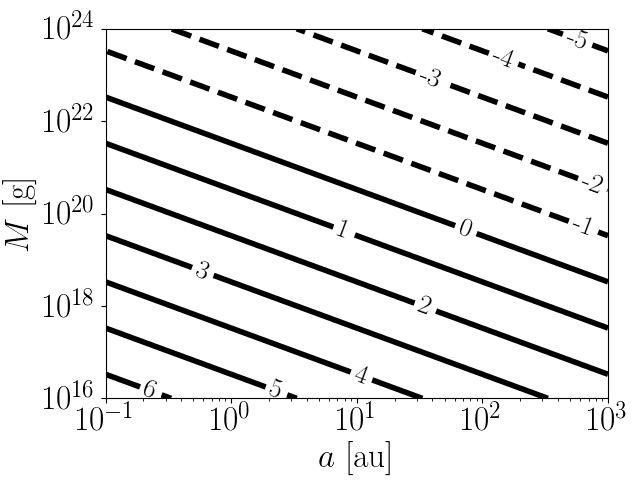}}
		\caption{Contour plot of lines of equal $\log_{10} \mathcal{Q}$ in separation $a$ as a function of disc mass $M$, given a fragment pericentre of $q/R_\odot = 1,0.6,0.2$ for panels (a), (b) and (c) respectively. The contours are normalized for fragment size $R=10\ \rm{m}$. The contour line labelled 0 means that a 10 m fragment is instantaneously captured, and the same outcome applies for the overlying dashed lines with negative numbers. Different fragment sizes should be changed according to $\mathcal{Q}\to\mathcal{Q}\times R/10\rm m$. Recall that in our model $M$ also provides an upper limit for the progenitor's mass.}
		\label{fig:1}
	\end{figure}
	
	Figure \ref{fig:1} shows contours of equal levels of $\log_{10} \mathcal{Q}$ as a function of the disc mass and fragment semi-major axes. The pericentre $q$ are indicated on top of each panel. The contours are normalized to $R=10\ \rm m$, while for other fragment sizes the scaling is simply $\mathcal{Q}\to \mathcal{Q} \times  R /10\rm m$. We see that the contours have identical slopes since $\mathcal{Q} \propto (Ma)^{-1}$. Larger separations and elevated disc masses both lower $\mathcal{Q}$, since the mass increases the energy extraction rate, while large separations decrease the orbital energy. A smaller pericentre distance $q$ results in a higher surface density at the intersection point and thus reduces $\mathcal{Q}$. Large fragments lose energy more slowly, since $\mathcal{Q} \propto R$.
	
	\begin{figure}
		\subfigure[$M=10^{19}$ g] {\label{fig:2a}\includegraphics[width=\columnwidth]{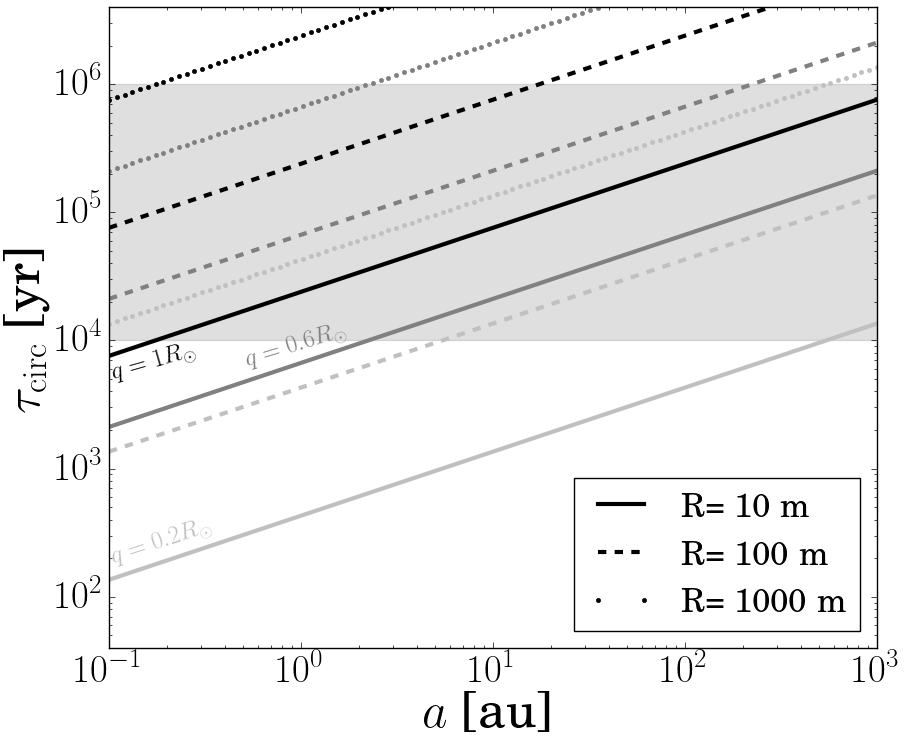}}
		\subfigure[$a=10$ AU] {\label{fig:2b}\includegraphics[width=\columnwidth]{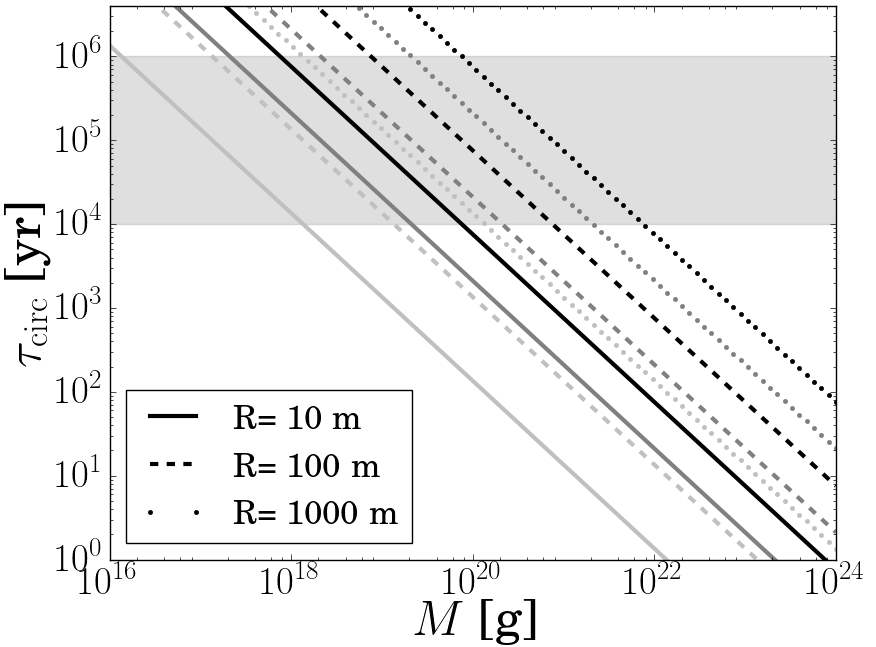}}
		\caption{Typical circularization time, $\tau_{\rm circ} = \mathcal{Q} P$ as a function of (a) the initial fragment semi-major axis (fixing $M$ at $10^{19}$ g, however for different $M$ change $\tau_{\rm circ}$ according to $\tau_{\rm circ}\to\tau_{\rm circ}\times 10^{19}{\rm g}/M$); and (b) the disc mass $M$ (fixing $a$ at 10 AU, however for different $a$ change $\tau_{\rm circ}$ according to $\tau_{\rm circ}\to\tau_{\rm circ}\times \sqrt{a/10{\rm AU}}$). Fragment size is indicated by line style, such that $R=$10/100/1000 m corresponds to solid/dashed/dotted lines respectively. Line colours correspond to fragment pericentre distances, lighter tones denoting shorter distances. The shaded area between $10^4-10^6$ yr is the empirically-based lifetime which might apply to the compact disc.}
		\label{fig:2}
	\end{figure}
	
	Figure \ref{fig:2} shows the circularization timescale. In Panel \subref{fig:2a} it is shown as a function of the separation when the compact disc mass is fixed at $M=10^{19}$ g. This timescale must be shorter or comparable to the compact disc lifetime (shaded area, based on \cite{GirvenEtAl-2012}). Note that these disc lifetimes are based on empirical evidence, rather than grounded by direct measurement or any physically motivated values. As we discuss later in Section \ref{SS:PartialCircularization}, the compact disc fate may sometimes be closely affected by its interaction with the tidal stream, so its lifetime can in fact be significantly shorter than that suggested by this empirical range. 
	
	The circularization timescale increases with the separation since $\mathcal{Q}P\propto a^{1/2}$. We see that small 10 m fragments (solid lines) are able to circularize within $\lesssim 10^6$ yr for any separation and any value of $q$. Intermediate 100 m sized fragments (dashed lines) also circularize within that time frame however they must intersect the disc at $q=0.6R_\odot$ or less, unless their separation is less than $10^2$ AU, in which case any value of $q$ is appropriate. Lastly, large 1000 m fragments (dotted lines) are considerably harder to circularize. They must be either of small separation or penetrate very deeply into the tidal sphere. Panel \subref{fig:2b} shows the circularization timescale as a function of the disc mass $M$ when $a$ is fixed at 10 AU. It is indicative of a wide range of fragment sizes and disc masses that circularize in propitious time. The minimum $M$ required at least for the smallest fragments is about $10^{16}$ g.
	
	\begin{figure}
		\subfigure[$q=1 R_\odot$] {\label{fig:3a}\includegraphics[width=\columnwidth]{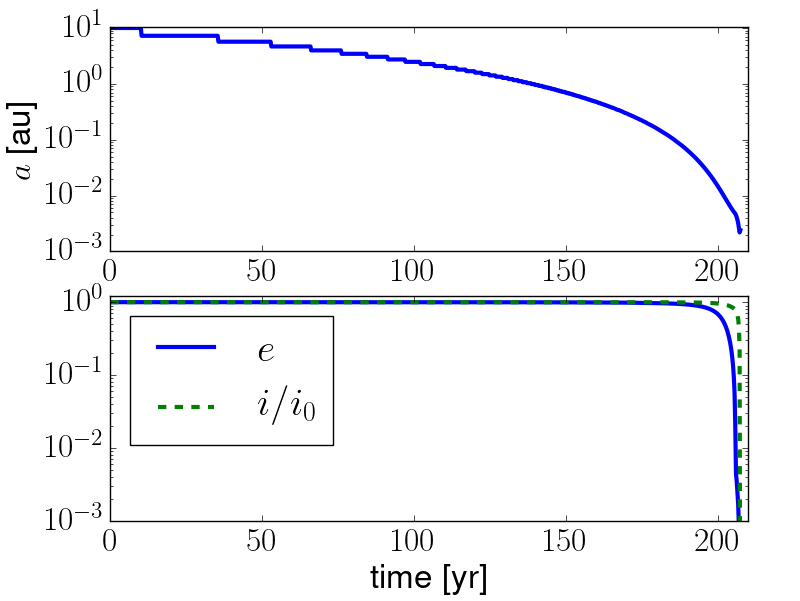}}
		\subfigure[$q=0.6 R_\odot$] {\label{fig:3b}\includegraphics[width=\columnwidth]{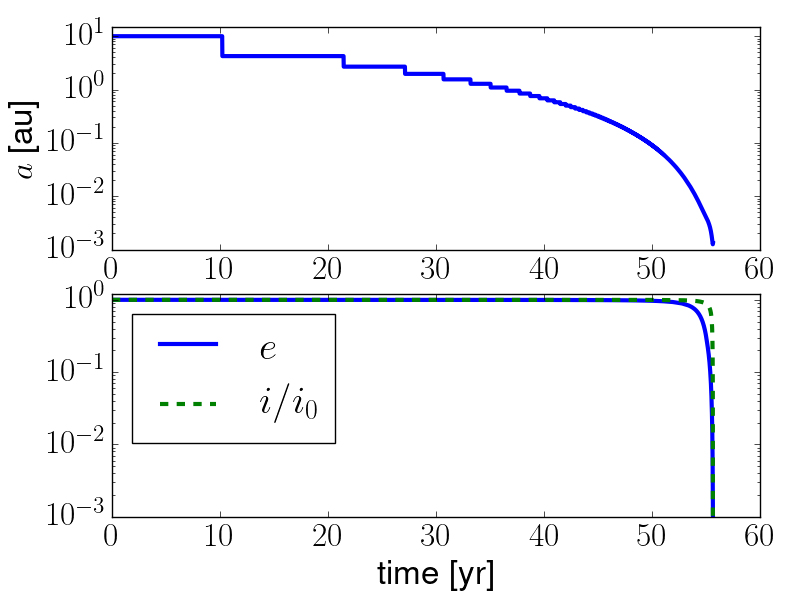}}
		\caption{Time evolution of the orbital elements of a fragment on initially eccentric orbit with $a=10 \rm au$ and initial inclination of $i_0=90^{\circ}$. The disc mass is $M = 10^{22}$ g. The pericentre is $q=1R_{\odot}$ in (a) and $q=0.6R_{\odot}$ in (b). The initial mean anomaly is $\mathcal{M}=3\pi/2$. }
		\label{fig:3}
	\end{figure}
	
	Finally, Figure \ref{fig:3} shows the time evolution by the numerical simulations described in Section \ref{SS:numeric}, of 10 m fragments initially separated by $a=10$ AU. The disc mass is fixed in the initial conditions of our code to be $10^{22}$ g. This choice is motivated by obtaining sufficiently short simulation runtime, of the order of $\sim$1 day. Panels \subref{fig:3a} and \subref{fig:3b} are similar except in having different choices of $q$. Each contain two subplots of the change in the semi-major axis $a$ and the corresponding change in the eccentricity $a$ and inclination $i$. Of course $e$ is initially $\sim$1 and $i$ is initially set to the maximum possible value of $90^{\circ}$ (face-on orientation).  
	
	We first discuss the damping timescale of $a$ versus $e$ and $i$. Since $\tau_{\rm circ}$ is defined above as the energy dissipation timescale, and since $E$ and $a$ are anti-correlated, $\tau_{\rm circ}$ is strictly speaking the migration timescale, or the rate of change of the semi-major axis. If we start with a very eccentric orbit of $e=1$ (or $a \gg q$), we eventually end up with a circular orbit of $e=0$ (or $a=q$). It is worth noting that the times for complete circularization and total elimination of the eccentricity are the same, however the eccentricity damping timescale, i.e the rate of change of the eccentricity with respect to its own magnitude, is different from $\tau_{\rm circ}$. This comes from the fact that the eccentricity often has initial values close to 1, and even after significant decrease of $a$ (say, of the order of itself) the eccentricity is still close to 1. Only when $a$ approaches $q$ does $e$ start to change significantly. As can be seen, $i$ and $e$ behave very similarly.
	
	The numerical simulations show that during the last phase of circularization fragments indeed become co-orbital with the compact disc and therefore are effectively embedded in it. Most importantly, Figure \ref{fig:3} and other simulation runs have all shown that there is a good agreement between the full numerical calculation and the analytical model. One expects the numerical simulations to reach full circularization faster than the analytical calculation since it is only an approximation and the exact drag changes as $a$ evolves. By comparing Figure \ref{fig:3} to Figure \ref{fig:2}\subref{fig:2b} using the same parameters, the timescales are clearly in agreement to order of magnitude, and the numerical circularization is indeed faster, as expected. Furthermore, in Panel \subref{fig:3b} it takes around $\sim 10$ orbits for $a$ to change in the order of itself, corresponding to the expected value of $\mathcal{Q}\approx 10$ from the analytical calculation. Conversely, in panel \subref{fig:3b}, $a$ changes significantly in one orbit, corresponding to the expected value of $\mathcal{Q} \approx 1$. The numerical simulations and analytical formalism are consistent.

	\section{Discussion}\label{S:discussion}
	
	
	\subsection{Gap formation and filling}\label{SS:GapFilling}
	In this section we quantitatively expand the discussion in Section \ref{SS:restrictions}, where it was suggested that each fragment effectively carves a column of dimension $R$ through the compact disc. In Appendix \ref{A:GapFillingTimescale} we derive from first principles a heuristic timescale for the diffusion filling timescale that will act to close the carved gap by local dust (ignoring the possibility of fully gaseous discs). We obtain the following equation:
	
	\begin{equation} \label{eq:tdif}
		\tau_{{\rm fill}}=\frac{3M R^2 q^{1/2}}{8h^2 \rho_{\rm d} r_{\rm out} r_{\rm d} (GM_{\rm WD})^{1/2}},
	\end{equation}
	
	\noindent where $M$ is the compact disc mass with an outer boundary $r_{\rm out}$, $q$ the pericentre distance, $h$ the compact disc scale height, $M_{\rm WD}$ the WD mass and $\rho_{\rm d}$ and $r_{\rm d}$ denote the density and radius of compact disc dust grains.

	When a gap forms by a fragment crossing the compact disc, our conceptual circularization model is valid as long as it fills up before another fragment overlaps the same region. All fragment crossings must encounter coincident dust rather than gaps, or else our model would be inconsistent with the calculation in Section \ref{SS:analytic}. We also recall from Sections \ref{SS:restrictions} and \ref{SS:PartialCircularization} that our only other condition is $M \gg M_{\rm progenitor}$.
	
	Our goal is now to consider the typical time between fragment crossings, and in particular the typical time between overlapping crossings. For simplicity, let us consider only same-size tidal fragments of dimension $R$, originating from a tidally disrupted progenitor of size $L$, such that all fragments occupy a similar semi-major axis $a$, as in the case of typical asteroid tidal disruptions (see Section \ref{SS:InitialFragDistances}). The intersection between the compact disc and each fragment occurs at the progenitor's pericentre distance $q$. Based on the MP20 study, all fragments intersect the compact disc within a relatively narrow cross-section approximately equal to the original progenitor's cross-section, i.e an area proportional to $L^2$. Clearly $L$ is many orders of magnitudes smaller than the dimension of the entire compact disc which is of order $R_\odot$, so the entire cross section is point-like in relation to the compact disc area. While this effective intersection point remains fixed in 3D space, the compact disc itself rotates around the WD.
	
	Additionally, since $R \ll L$, the probability of successive fragments in the tidal stream to overlap exactly at the same intersection point is of the order of $R^2/L^2$, assuming a uniform spatial distribution inside the stream. The total number of fragments in the stream $N$ is given by the mass in the stream divided by the mass of each fragment, or $N=L^3/R^3$ (constant $\rho$ assumed). Thus the typical time between crossings in this simplified setup is the fragment orbital period $\sqrt{a^3/GM_{\rm WD}}$ divided by N:
	
	\begin{equation}\label{eq:CrossingTimescale}
	\tau_{\rm cross} = \sqrt{\frac{a^3}{GM_{\rm WD}}}\frac{R^3}{L^3}.
	\end{equation}
	
	However considering the probability to cross in the same exact point, the typical time between overlapping crossings is $\sqrt{a^3/GM_{\rm WD}}$ divided by $L/R$:
	
	\begin{equation}\label{eq:OverlapTimescale}
	\tau_{\rm overlap} = \sqrt{\frac{a^3}{GM_{\rm WD}}}\frac{R}{L}.
	\end{equation}
	
	If the gap is unable to rotate in time to avoid an overlapping crossing, the model is invalid. The compact disc rotation period at a radial distance of $q$ from the WD is (using Kepler's law) $P=2 \pi q^{3/2} (GM_{\rm WD})^{-1/2}$. The disc must rotate a small distance of approximately $R$ in order to avoid a second hit by an overlapping crossing in the same point. The fractional rotation time is $R/2 \pi q \times P$. We equate this with $\tau_{\rm overlap}$ to constrain the semi-major $a$. The resulting expression is $a<q^{1/3}L^{2/3}$. Since the progenitor size $L$ obviously cannot be larger than the size of the disc, $a$ cannot exceed the pericentre. The conclusion is that $\tau_{\rm overlap}$ must be much larger than the fractional rotation time. Overlapping crossings are sufficiently temporally separated that the disc always rotates away.
	
	The gap thus has to complete \emph{at least} one full orbit around the WD before it has some probability to intercept an additional fragment. Therefore, one rotation period of the compact disc is a minimal requirement for the time interval between gap fillings. Based on this, $\tau_{\rm fill}$ from Equation \ref{eq:tdif} must be smaller than $P$, and we can rearrange this equation to obtain a strong limit on the maximum disc mass allowed for the model to work:

	\begin{equation}\label{eq:Mcriteria}
	\begin{split}
	M \lessapprox 2 \times 10^{29} \: \mathrm{g} \:   \left( \frac{R}{10~\rm{m}} \right)^{-2} \left( \frac{q}{R_\odot} \right) \left( \frac{h}{1000~\rm{km}} \right)^{2} \\
	\left( \frac{\rho_{\rm d}}{3~\rm{g} \times \rm{cm}^{-3}} \right) \left( \frac{r_{\rm d}}{1~\mu\rm{m}} \right). \\
	\end{split}
	\end{equation}
	%

	We take the fiducial values $\rho_{\rm d}=3~\rm{g} \times \rm{cm}^{-3}$ and $r_{\rm d}=1~\mu\rm{m}$. We then consider extremum values $R=1000$ m and $q=0.2 R_\odot$ to obtain $M \lessapprox 4 \times 10^{24}~\rm{g} \times (h / 1000~\rm{km})^2$. Based on our compact disc mass range from Section \ref{SS:Discs} ($10^{16}~\rm{g}<M<10^{24}~\rm{g}$), Equation \ref{eq:Mcriteria} implies that our model is robust even for 1000 m gaps, unless $h$ is considerably smaller than 1000 km. Since we consider fragments that follow a power law size distribution, most gaps will in fact be of dimension $R=10$ m, in which case the model is robust for any $M$ and $h$ down to 1-10 km. 
	
	The scale height $h$ is observationally an unconstrained parameter. Nevertheless, the \cite{KenyonBromley-2017a,KenyonBromley-2017b} theoretical studies investigate the evolution of compact dusty discs around WDs. It is shown that if the debris inside the compact disc are of infinite strength, collisional evolution leads to a scale height as small as just one or two times the size of the largest debris. However, in the more plausible case where debris suffer catastrophic disruptions, the scale height might be reduced by a factor of at most 2. The authors then show that the scale height will generally remain in the order of $\sim 10^3\ \rm km$, corresponding to $f=0.002-0.004$. Given these arguments we conclude that $h$ is very likely compatible with our model being robust, and the gaps fill quickly prior to completing one orbital period (of order $\sim$ few hours). Empirically, in WD 1145+017 \citep{VanderburgEtAl-2015} $h$ must be large, otherwise eclipses of the WD by orbiting solids would not be observed. However this is merely one example. We also show in Appendix \ref{A:GravInstabilities} that the $h$ range considered in this work is safely larger than the limit case that leads to gravitational instability in the compact disc.

	\subsection{A ram-pressure catastrophe?}\label{SS:CatastrophicOutcome}
	Could the response of the fragment to passing through the compact disc be catastrophic? As discussed in Section \ref{S:model}, gas or dust collectively lead to a large pressure difference between the front and the back of the fragment. This so called ram pressure (or dynamical pressure) may be seen as equivalent to the energy density, which can be worked out from Equation \ref{eq:de}.
	
	\begin{equation}
	P_{\rm ram} =  \rho_{\rm disc} v_{\rm rel}^2 = \frac{(C\sim 1)\Sigma (q) v_{\rm rel}^2}{h}.  \label{eq:Pram}
	\end{equation}
	
	As previously mentioned the scale height $h$ is a quantity that is poorly constrained by physical models, ranging from one or two dust radii to a few thousand km \citep{KenyonBromley-2017a}. Because the relative velocities at pericentre are so large (few $\sim 10^2$ km/s), we check if the resulting ram pressure can exceed the threshold for compressive fracture. Measurements of meteoroids in the Earth's atmosphere yield compressive strengths in the broad range of 1-500 Mpa \citep{Petrovic2002, Popova2011, Podolak2015}. Even in the limit of the smallest compressive strength in this range (1 MPa), the scale height must be considerably less than $\sim 10^3$ km and the disc mass must exceed $ \sim 10^{23}$ g for $P_{\rm ram}$ to be important. In reality, the actual compressive strength of large cohesive fragments inside the Roche limit could be in the neighbourhood of 500 Mpa, otherwise they are not likely to withstand the tidal forces and remain cohesive monolithic objects (for example, the assumed strength in \cite{BrownEtAl-2017} is 1 GPa). Likewise, disc masses approaching $\sim 10^{23}-10^{24}$ g should be extremely rare, as already discussed in detail in Section \ref{SS:Discs}. Therefore, plugging in a more plausible disc mass of $M=10^{21}$ g, we conclude that the compact disc scale height $h$ has to be on the order of meters, same as our smallest fragments, in order for the ram pressure to be important. Ram-pressure is thus deemed as an unlikely mechanism to bring about catastrophic outcomes in strong cohesive or monolithic fragments.
	
	However, for weak comet-like material, it is an open question to what extent fragments are susceptible to catastrophically disintegrate. As an example consider the compressive strength recently measured for the surface of comet 67P/Churyumov–Gerasimenko, an astonishingly low compressive strength of merely 12 Pa \citep{ORourkeEtAl-2020}. Such weak material will certainly not survive. Nevertheless, the surface strength mentioned above probably does not reflect on the bulk strength of 67P/Churyumov–Gerasimenko. Other estimates have yielded compressive strengths ranging from 1 kPa \citep{BieleEtAl-2015} to 2-4 MPa \citep{SpohnEtAl-2015}, so the effective strength is probably related to depth variations. 
	
	Given the above discussion, it remains to be determined if comet-like fragments are resistant to ram-pressure breakup, yet cohesive rocky fragments are in all likelihood strong enough and will not be affected. Independent of ram-pressure, fragment erosion is likely, and its effect is discussed in the following section.
	
	\subsection{Gas production and fragment erosion}\label{SS:GasProduction}
	How gas is actually being produced in WD discs is one of the fundamental open questions in the field, although as previously discussed, gas is rarely observed and is much more infrequent than dust \citep{ManserEtAl-2020}, although the fraction of gaseous discs has recently increased considerably. We discuss several possible mechanisms. First, gas may be produced by \textbf{sublimation of dust} at a sufficiently close distance from the WD (see for example \cite{MetzgerEtAl-2012} and references therein). For this mechanism, the presence and radial extent of the gas should clearly correlate to some degree with the cooling age (hence luminosity) of the WD. If collisions among grains at velocities of at least few $10^{-1}-10^0$ km $\times$ s$^{-1}$ occur, they would lead to shattering \citep{TielensEtAl-1994}, thereby reduce the grain sizes and enhance the effectiveness of sublimation. The classical dust sublimation 'radius' is in fact not a radius but an annulus of different grain-size-dependent sublimation rates. Additionally, grains of volatile materials are more prone to sublimation and if they are introduced for any reason, even within an optically thick compact disc, they would certainly sublimate at greater distances than the dust.
	
	A second possibility is gas being produced via \textbf{grain-grain collisional vaporization}. In this process the collisions between projectile grains and similar-sized or else much larger target grains result in compression shocks which lead to increase in pressure and temperature, and subsequently phase transformations in the material \citep{AhrensOkeefe-1972}. Driven essentially by the kinetic energy of the impacting bodies, studies have shown that partial vaporization can be incepted when the internal energy behind the shock is about 1-2 times the material surface binding energy, which is the energy necessary in order to remove an atom from the top surface layer in vacuum \citep{KudriavtsevEtAl-2005}. Full vaporization is possible at about 5 times the material surface binding energy (see \cite{TielensEtAl-1994} and discussion therein). The corresponding collision velocities required of similar-sized silicate grains is of the order of a few $10^1$ km $\times$ s$^{-1}$ for partial vaporization, while full vaporization requires $\sim 10^2$ km $\times$ s$^{-1}$. If not similar-sized, but instead having a lower projectile:target mass ratio (e.g, 0.1 in Figure 9 of \cite{TielensEtAl-1994}), the smaller projectile is more likely to fully vaporize while the vaporized fraction in the target is reduced compared to same-size collisions. Full vaporization in the target then requires typical collision velocities a few times larger than those mentioned in the case of similar-sized grains. 
	
	Tidally disrupted fragments are however larger than mere dust grains (say, tens or hundreds of meters versus mere micron-sized projectile particles). While collisional vaporization also occurs when the target is infinitely larger than the projectile, in such cases the vaporized mass in the target is negligible compared to the mass lost to shattered dust particles, excavated from the underlying (i.e deeper) shocked surface layers \citep{TielensEtAl-1994}. The term shattering here is equivalent to cratering, and it is very important for fragment erosion. An impacting dust particle hits the fragment, producing a crater on the surface. As a rule of thumb craters have about 10-20 the linear dimension of the impactor, hence the excavated volume (or mass) is some $\sim 10^3$ times larger than that of the impacting grain. However, as we show in Appendix \ref{A:ShatteringYield}, the dusty ejecta from the fragment's surface collides with dust grains in the compact disc at high velocities, causing grain-grain vaporization. This quickly forms a blanket of vaporizing material that effectively prevents almost all subsequent compact disc projectile dust from reaching the fragment's surface. Instead of a high yield of the order of $10^3$ we have an effective yield which is shown to be $Y_{\rm shat}\approx$1.
	
	An influx of tidally disrupted fragments, infinitely larger than mere dust grains, could therefore interact with a pre-existing compact disc. Given the previously mentioned tidal radius of typical WDs, the parabolic pericentre Keplerian velocity of an extremely eccentric ring of fragments would be several times $\sim 10^2$ km $\times$ s$^{-1}$, while the circular (or near-circular) Keplerian velocity of a pre-existing compact disc is a factor $\sqrt{2}$ lower for the same distance. Hence even if the fragments are exactly co-orbital with the compact disc, the relative collision velocities are extremely high. Additionally, any inclination relative to the pre-existing compact disc considerably raises the relative collision velocities. For a more dispersive tidal stream, as in the MP20 study, fragments of different eccentricities converge at periecntre. In this case, even the velocity dispersion within the stream itself is significant (however we do not discuss mutual fragment collisions in this study, only their interaction with the compact disc). Therefore, given the previously mentioned shattering and vaporization threshold velocities, it is easy to speculate in this scenario that conditions are highly appropriate for collisional gas formation, driven by vaporization of two components: pre-existing compact disc dust and eroded dust originating from the shattered fragments.
	
	A third possibility is gas generation via \textbf{sputtering}, a process in which energetic gas ions collide with solid material. Colliding ions set off collision cascades among the atoms inside the target material, and some atoms may recoil back toward the surface of the target. If indeed a collision cascade reaches the surface of the target, and its remaining energy is greater than the target's surface binding energy, an atom is ejected \citep{BehrischEckstein-2007}. Thus, sputtering requires the presence of pre-existing gas intersecting solid material. The sputtering yield (i.e, atoms sputtered per gas ion) depends on the deposited energy per unit length to the surface binding energy. Hence, not merely the target material is important, but also the collision velocity and mass of the gas ions. For typical silicate material (of both the gas and solid targets) and given the likely velocities as argued in the previous paragraph, the sputtering yield $Y_{sput}$ is of the order of 0.1-1, and approximately an order of magnitude larger if the material is $H_2O$ (see figure 10 in \cite{TielensEtAl-1994}). A similar sputtering yield of 0.1 was adopted in the study of \cite{Jura-2008} for silicates. The only pre-requisite is coincident gas to trigger the process.
	
	A forth and final possibility might be related to a recent discovery of a planetesimal held together by its own internal strength, embedded inside the well studied debris disc of SDSS J1228+1040 \citep{ManserEtAl-2019}. These authors have used short-cadence spectroscopy in order to search for signs of random variations in the gaseous emission of the WD, which they hypothesized would be produced by random gas-generating collisions among dust particles. Instead they found periodic variations, and concluded that the variability in emissions must result from an excited cloud of gas trailing a cohesive object whose size is unconstrained but may be up to a few hundred km in radius, and having a two-hour orbital period. Or else, they hypothesized that the planetesimal itself could be producing the gas if its orbit is close enough to the WD to trigger surface sublimation. In the first case, however, it is not yet clear what sort of interaction between the planetesimal and the disc could generate the gas.
	
	The relative velocity of an eccentric fragment with eccentricity $e$ embedded within a circular (assumed for simplicity) compact disc is given by $(\sqrt{1+e}-1)v_{\rm circ}$. Even $e$ as low as 0.1-0.2 should induce sufficient relative velocities (a few dozen km $\times$ s$^{-1}$) for partial collisional vaporization of dust and also erode an embedded fragment through shattering to replenish dust. Once the relative velocity is small enough, none of the above is expected. The SDSS J1228+1040 embedded object's semi-major axis is however modelled as 0.73 $R_{\odot}$, without any appreciable change between observations separated by almost a year \citep{ManserEtAl-2019}. Therefore it does not seem likely that in this case a small primordial eccentricity is responsible for producing the gas. I.e, the planetesimal is fully embedded in the compact disc rather than still in the process of reducing its eccentricity to match that of the compact disc. The precise details of how the variable gas emission is generated thus requires further investigation. Nevertheless, it is clear that at least during the onset of fragment entrapment in the compact disc, continuous and widespread gas generation is expected. Also, the case of SDSS J1228+1040 emphasizes that additional physics might be important.
	
	For completion, we also note that for a gaseous compact disc, aeolian wind erosion is also capable of eroding the surface of fragments for typical relative velocities of a few dozen m $\times$ s$^{-1}$, which corresponds to extremely low $e$ in the range 0.0001-0.0002. Wind erosion is however a purely mechanical process, not unlike the erosion of sand dunes. It applies to weakly consolidated surfaces, and was initially invoked for protoplanetary discs \citep{RoznerEtAl-2020a,GrishinEtAl-2020b}. Recent study of wind erosion by \cite{RoznerEtAl-2020b} assumes a-priori that fragments with the appropriate mechanical properties reside in WD disc. In our study however the fragments must first \emph{become} embedded in the compact disc by circularization, hence they inescapably first interact with the disc at much higher velocities. In turn, if they are weakly consolidated, it is unclear that they would survive those interactions or the ram pressure analysis discussed in Section \ref{SS:CatastrophicOutcome}.
	
	A conceivable picture finally emerges from the above arguments. At the initial stages of tidal debris formation, following a tidal disruption event of a planetary object, an influx of tidally disrupted fragments should interact with a pre-existing compact disc and generate gas via sputtering \citep{Jura-2008} if fragments meet obstructing gas along their path. In addition, any coincident dust grains in the pre-existing compact disc would be \emph{entirely} collisionally vaporized, and so would similar-sized or slightly larger grains in the tidal stream itself. Larger tidal fragments (of sizes ranging between $10^1-10^3$ m) will collisionally self-erode mainly through shattering, however the secondary effect is likewise grain-grain collisions, again leading to gas production. Hence, large fragment self-erosion would replenish the compact disc in new dust and gas, however not instantaneously, since the arguments in Section \ref{SS:analytic} make it clear that material in the compact disc initially attains the same velocity as the fragments themselves.		
	
	Finally, if some fragments bear volatiles or super-volatiles, all of the above-mentioned mechanisms will be enhanced. Namely, the sputtering yield would increase by at least an order of magnitude, and any collisionally shattered grains would have an increased chance of being sublimated even if shadowed by dust inside the opaque compact disc. Fragments whose orbital energy has been sufficiently dissipated by the interactions with the compact disc become trapped in it. Then there is at least a short phase during which relative velocities are sufficient to continuously rather than impulsively generate more widespread gas. Arguments presented in the MP20 and \cite{VerasEtAl-2014} studies indeed favour such an influx of fragments at the earlier stages of tidal debris disc formation. Hence, we predict gas formation to be an inescapable part our model. In the following section we continue discussing the conditions for having a steady gas component in the compact disc.
	
	\subsection{Observational consequences}\label{SS:ObservationalConsequences}
	Until now certain studies (e.g., \cite{GrishinVeras-2019} and \cite{OConnorLai-2020}) have assumed that a pre-existing compact disc exists and that it has a gaseous component, however without specifying why. In the previous section we have shown how a passing tidal fragment can interact with a compact disc to generate gas, and we shall now further discuss under what conditions the gas production can be ongoing. This will be the first observational consequence of our model, and will depend on the balance between gas production and condensation. The second potential consequence of our model is the formation of a dust halo, leading to infrared variability.
	
	Consider again the various gas forming mechanisms, presented in Section \ref{SS:GasProduction}. Even if we completely neglect gas production by grain-grain collisional vaporization, sputtering or sublimation of volatiles, collisional vaporization of pre-existing dust grains in the compact disc (by a crossing fragment) would convert an entire column of dust with volume $hR^2$ and mass $\Sigma R^2$ to gas. Additionally, more dust would be produced from self-erosion of the fragment by shattering. According to the discussion in Section \ref{SS:restrictions}, all this material should initially obtain a velocity of similar magnitude to that of the fragment. It is reasonable to assume that not all of this material will attain exactly the fragment trajectory. More likely, at least some part of it might scatter in different directions by the interaction. This material would attain different orbits than the stream. Neglecting the details of this chaotic scatter, we can simply assume that the interaction leads to various concentrations of gas and dust around the WD, forming a sort of halo, in addition to material accompanying the fragment.
	
	Let us now consider a tidal fragment following the interaction with the compact disc. We have just outlined that as it emerges from the compact disc it is surrounded by a cloud of gas and dust of mass $\sim \Sigma R^2$, initially travelling along the original fragment trajectory. However, if indeed even a thin halo of material exists, gas and small dust grains would quickly decelerate to match the velocity of the halo whereas the infinitely larger (relatively speaking) fragment would continue uninterrupted. What will be the probable fate of such gas and dust? First we note that the temperatures above and below an opaque compact dust disc are necessarily much higher than inside the compact disc due to the absence of shadowing (see e.g. \cite{MelisEtAl-2010}), and therefore even refractory dust should sublimate on some timescale, unless sufficiently far from the WD (also depending on its cooling age). Whether dust or gas, \cite{FarihiEtAl-2018} estimate that any material should be quickly re-assimilated into the underlying compact disc, unless radially extending beyond the disc extremities (inner or outer edge). This is because if their orbits cross that of the compact disc, the latter would damp relative dust inclinations and/or act to re-condense gas on orbital timescales. In what follows we discuss these two options, namely, halo material internal to the compact disc extremities, which could lead to detectable ongoing gas, and halo material outside the outer edge of the disc which could lead to infrared variability.
	
	\subsubsection{Ongoing gas production}\label{SSS:OngoingGas}
	As we have previously described, a constant flux of fragments necessarily produces gas. However the gas might only be present at quasi steady state if the rate of production exceeds the rate of gas condensation. Let us then consider the condensation timescale of the gas. In the study of \cite{MetzgerEtAl-2012}, the authors state that it is not clear how gas can exist simultaneously in stable phase equilibrium with adjacent dust. The compact dust disc is believed to be optically thick and the temperature of the solids is necessarily below the sublimation temperature for all but the innermost annulus of dust (otherwise the entire compact disc would all have been gaseous). If hotter gas exists at any time, atoms of gas should stick upon colliding with the surface of dust particles, resulting in condensation. The probability of sticking is given by the so-called sticking coefficient $\alpha$ \citep{Leitch-DevlinWilliams-1985}. The condensation timescale $\tau_{\rm cond}$ is shown to be $\sim P / \alpha$, where $P$ is the orbital time of the gaseous component at separation $q$. We thus have:
	
	\begin{equation}\label{eq:CondensationTimescale}
	\tau_{\rm cond} = \alpha^{-1}\sqrt{\frac{q^3}{GM_{\rm WD}}}.
	\end{equation}
	
	In the study of \cite{MetzgerEtAl-2012}, Equation \ref{eq:CondensationTimescale} constituted as a serious problem since in the absence of steady gas formation in their model, any existing gas should condense in typical timescales of a few to a few hundreds of orbital periods, depending on how small $\alpha$ is (one orbital period is $\sim$3-4 h at $r=R_{\odot}$). In our model this problem is greatly alleviated by the fact that we do have a steady mechanism that indeed forms gas continuously. As shown by MP20 and other studies, during the early phase of tidal disruption and debris disc formation, an influx of eccentric tidal fragments ensues. For continuous gas we require $\tau_{\rm cross}<\tau_{\rm cond}$, otherwise gas is only produced intermittently. We use Equations \ref{eq:CrossingTimescale} and \ref{eq:CondensationTimescale} to obtain a condition on the semi-major axis $a$ such that
	
	\begin{equation} \label{eq:gasProduction}
	a \lessapprox 5000~\rm{AU} \left( \frac{q}{R_\odot} \right) \left( \frac{L}{10~\rm{km}} \right)^{2} \left( \frac{R}{10~\rm{m}} \right)^{-2},
	\end{equation}
	
	\noindent where we assume $\alpha$ is approximately of order unity \citep{Leitch-DevlinWilliams-1985}. Taking fiducial values of $q=R_\odot$ and $L=10$ km (an asteroid), we see that gas production is continuous for 10 m fragments virtually irrespective of the separation in the range $a<10^3$ AU. On the other hand, 1000 m fragments require $a<0.5$ AU which almost never happens in our model. 
	
	Equation \ref{eq:gasProduction} therefore provides an interesting prediction. During the initial stages of a tidal disruption event, the tidal stream is still abundant with small fragments. These however circularize much faster than the large fragments, as clearly evident in Figures \ref{fig:1} and \ref{fig:2}. After the small fragments are embedded in the compact disc, the number of remaining large fragments is much lower (given an initial power-law fragment size distribution) and therefore while gas is still being produced, it happens discontinuously and the gas quickly condenses. Therefore, gas is probably much easier to detect during the initial phases following tidal disruptions, and much harder to detect as tidal streams mature and evolve. This prediction is in agreement with the observation that fewer discs are observed to have a gas component compared to the number of discs that have dust \citep{ManserEtAl-2020}, if the progenitor is assumed to be asteroid-sized. It is also seemingly compatible with the case of SDSS J1617+1620, where strong double-peaked CaII emission lines appeared in 2008 and monotonically faded over a period of a few years \citep{WilsonEtAl-2014}. In our model, a sufficiently small asteroid progenitor might produce this pattern, having only a small mass in the ensuing tidal stream, which evolves rapidly. We note to contrast, that large progenitors have large $L$, in addition to tidal fragments that typically have small $a$ and a wide spread in orbital energies. This requires a different derivation for $\tau_{\rm cross}$. Also, their mass has a larger probability to exceed that of the pre-existing compact disc, suggesting that at the end of a partial circularization phase, gas production (if it occurs) must ensue from other processes and no longer through interacting with a pre-existing compact disc.
	
	We also note that gas production in the early stages following a tidal disruption event is contributed by self-erosion of small fragments, as will be discussed in more details in Section \ref{SS:ErosionVsCircularization}.
	
	
	Finally we note that another potential consequence could be variability in the UV. Since most circumstellar absorption lines are concentrated in the UV \citep{XuEtAl-2019}, variable concentration of halo gas might yield UV variability and this could be checked by performing UV photometry of known gaseous discs.
	
	\subsubsection{Detection of infrared variability}\label{SSS:InfraredVariability}
	
	\begin{figure*}
		\includegraphics[width= \textwidth]{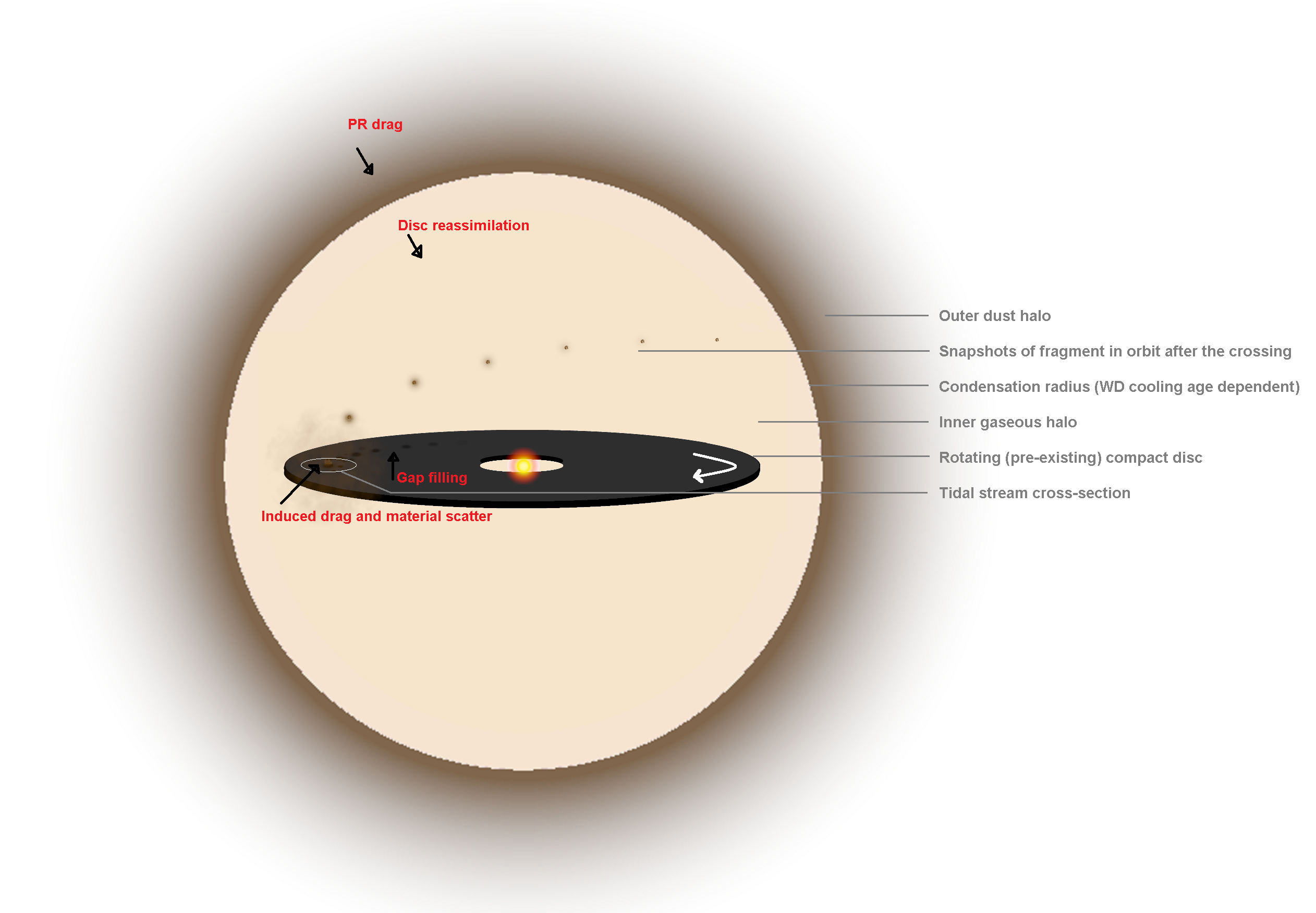}
		\caption{A schematic diagram (not to scale) depicting the interaction of tidal fragments with the pre-existing compact disc, with possible observational consequences. An influx of fragments from a tidal stream is collisionally interacting with the compact disc. During each crossing a gap forms in the rotating compact disc, whose size depends on the fragment size. The fragment scatters material in the collision and continues on its new trajectory after drag-induced orbital dissipation. As the compact disc rotates gaps are filled through diffusion, however large gaps fill more slowly. Fragments are accompanied by a trailing cloud of material (from compact disc and self-erosion). These clouds experience sublimation/condensation depending on radial distance from the WD. They decelerate through interacting with an inner halo of gas and outer halo of dust. High concentrations of gas in the inner halo can exist if the influx of tidal fragments exceeds the rate of condensation onto the compact disc. Dust in the outer halo infalls through PR drag and may produce infrared variability. Eventually migration inwards of the compact disc's outer edge results in re-assimilation.}
		\label{fig:4}
	\end{figure*}
	
	Assuming that some small fraction of material is scattered beyond the compact disc outer edge (hereafter the 'outer halo'), what is the significance and possible implications of such material? As previously stated, the temperatures above and below an opaque compact dust disc are higher than in the shadowed internal parts (see e.g. \cite{Jura-2003} and \cite{MelisEtAl-2010}). Depending on the exact WD cooling age, scattered and unshielded refractory dust might sublimate on some timescale, unless the dust grains are too large. However, material reaching far enough from the WD would necessarily condense to form dust. Regardless of WD cooling age, we thus speculate that an outer halo exists, far enough from the WD that it may be composed primarily of optically thin dust grains. Some of this dust could initially scatter to these wide orbits around the WD during the collisions between fragments and the compact disc. More dust is perhaps accumulated in this outer halo through interacting with the tidal fragments (and associated clouds of material around them) that pass through it. A third option is dust produced by large (in our model larger than about 1000 m), tidally dissecting fragments. Although the dissected sub-fragments are much larger than mere dust \citep{Rafikov-2018}, a small fraction of dust may be released, adding to the outer halo dust. In this work we do not provide a calculation of the exact amount of dust to reach the outer halo, since this calculation is beyond the scope of our current paper. This requires a more specific and detailed model which should be left to future dedicated studies. We however discuss the fate of such material if indeed it exists.
	
	Extending beyond the outer edge of the compact disc, rapid damping by direct contact with the compact disc would no longer be possible. However, PR drag would cause the dust grains to drift inward. At some point they might sublimate, depending on their size and the WD cooling age. When the gas/dust eventually falls below the outer edge of the compact disc it would be re-assimilated in it as discussed in Section \ref{SSS:OngoingGas}. The dominant timescale in the above picture is $\tau_{\rm PR}$, the timescale of radial drift via PR drag. From \cite{KenyonBromley-2017a} we have:
	
	\begin{equation}\label{eq:PRtimescale}
	\tau_{\rm PR} \cong 5 {\rm yr} \left( \frac{d}{1 \mu m} \right) \left( \frac{a}{R_\odot} \right)^2 \left( \frac{10^{-2}L_\odot}{L_{\rm WD}} \right),
	\end{equation}
	
	\noindent where $r_{\rm d}$ is the grain size, $a$ its semi-major axis (assuming small eccentricity) and $L_{\rm WD}$ the WD luminosity. The radial drift timescale is in the order of years, however the grains must only travel a fraction of the distance to the WD (they cease to be optically thin dust when they fall below their sublimation radius or reach the outer edge of the compact disc). Additionally, grains of various sizes drift at different speeds and therefore collisional damping might also expedite the inward radial drift. Overall the required timescale might even be weeks or months. 
	
	Interestingly, \cite{FarihiEtAl-2018}, \cite{SwanEtAl-2019}, \cite{WangEtAl-2019} and more recently \cite{SwanEtAl-2020} have shown evidence for widespread infrared variability which is found across a large population of stars. Not only that, increased variability correlates with stars which also feature emitting gas. The timescale of highest variability is months rather than hours or days, noting however that very few observations in the sample are spaced less than a week apart. \cite{FarihiEtAl-2018} and \cite{SwanEtAl-2020} interpret these findings as related to collisions, without providing a detailed model. However, as outlined in Section \ref{SSS:OngoingGas}, if the optically thin dust is internal to the compact disc extremities, these authors cannot easily accommodate their proposed collisions with the long variability timescales observed.
	
	If however an outer halo exists, the timescales are more compatible. Any infrared variability on timescales longer than hours (approximately corresponding to re-assimilation in the compact disc) should be governed by how frequently large deposits of dust reach the outer halo. If the time between such deposits is longer than the aforementioned PR drag timescale then the PR drag timescale would be subdominant. If the opposite is true, then variability could be a signature of the fragment size distribution, or perhaps a sign of clumping of fragments in the tidal stream (which usually correlates with an early phase in the evolution of the tidal stream). More accurate and detailed models are required in order to develop this idea further.
	
	For completeness we also note that in a recent 3 yr photometric near-infrared survey of 34 WDs in the J, H and especially the $\sim 2 \mu $m K band which is more sensitive to dust in comparison, a largely stable flux with no significant variability has been reported \citep{RogersEtAl-2020}. This contrasts with previous K-band variability reports ranging between 13\% \citep{XuEtAl-2018} and 18.5\% \citep{XuJura-2014} in two respective WDs. Of course the \textit{Spitzer} $3.6 \mu $m and $4.5 \mu $m bands which were mentioned earlier \citep{FarihiEtAl-2018,SwanEtAl-2019,WangEtAl-2019,SwanEtAl-2020} are more sensitive to the dust variability and thus we interpret our model based on these findings.
	
	Finally, we wish to emphasize that the idea raised in this section is essentially \emph{independent} of whether or not gas production is ongoing or intermittent. However, if gas production is ongoing it probably means that there is a much higher concentration of gas and perhaps dust inside the inner halo (i.e the part of the halo having a radial distance smaller than $r_{\rm out}$). In turn fragments travelling through the inner halo have a greater chance of scattering material from the inner to the outer halo, hence one might expect increased variability. That is in addition to the fact that the fragment crossings are more numerous at earlier times implying more material is scattered. \cite{SwanEtAl-2020} have indeed shown that discs that also have gas in emission feature higher variability.
	
	Figure \ref{fig:4} visually captures the ideas proposed in Sections \ref{SSS:OngoingGas} and \ref{SSS:InfraredVariability}. Interestingly, somewhat similar ideas regarding halo structures have already been proposed based on the merit of fitting the observational evidence (e.g., see section 3.3 in \cite{Farihi-2016}).
	
	\subsection{Forming the previous compact disc}\label{SS:FirstDisc}
	A caveat of the model is obviously the requirement for a pre-existing compact disc. In other words, in the framework of our model it is impossible to shrink an eccentric tidal stream generated by a recent tidal disruption event without first having a compact disc around the WD with which it can interact. Within the confines of our model we nevertheless hypothesize that the recent progenitor that generated the tidal stream eventually adds its own mass and replenishes the exiting compact disc. Under certain conditions, which are discussed in the next Section (\ref{SS:ComparableMasses}), this could allow for longevity of the compact disc, ensuring its survival over multiple tidal disruption events. Alternatively, the compact disc may also be dispersed under other conditions, which raises the questions - what other processes can initially generate the pre-existing compact disc for our model to work?
	
	We would argue that this is not a 'chicken and egg' type of problem, since the compact disc can probably form without dust-assisted circularization, only on longer timescales or else assuming a different composition. In Section \ref{S:Intro} we have already mentioned some of the other potential mechanisms that may enable the formation of a compact disc, and we would repeat them briefly: (1) the Yarkovsky effect is in theory capable of affecting tidal fragments of the size we have explored \citep{VerasEtAl-2015}; (2) if collisional cascade may break initially large fragments to mere dust \citep{WyattEtAl-2011} then the disc may shrink via PR drag \citep{VerasEtAl-2014}; (3) perhaps our favourite alternative is a tidal stream consisting of fragments so highly rich in volatile/super-volatile materials that it deposits a large fraction of its mass in sublimated gas during its preliminary or secondary flyby around the WD. This gas would trigger the onset of circularization via our model, initially through gas drag only, without the need for other models. Nor does it require long timescales in order to operate.
	
	\subsection{Implications of partial circularization}\label{SS:ComparableMasses}
	We recall from Sections \ref{SS:restrictions} and \ref{SS:PartialCircularization} that when the tidal stream progenitor mass is comparable or larger than the compact disc mass, i.e $M \lessapprox M_{\rm progenitor}$, we have partial circularization. The compact disc is dispersed and we are left with a partially circularized tidal stream + scattered optically thin material in close WD orbit. As we shall see in the following section, fragment erosion in that case is negligible.
	
	If the the tidal stream progenitors are equal-sized, what would be the model implications? Suppose the time interval between planetesimal injections (which give rise to tidal disruptions and subsequent tidal streams) is shorter than time it takes the compact disc to form without a pre-existing disc (i.e, through a combination of Yarkovsky, catastrophic collisions, PR drag or some other processes). In such a case, we have no mechanism to form the canonical compact disc in the first place, since if it would form - it would be dispersed. Accretion onto the WD would be mainly controlled by PR drag. If instead some WD systems are configured to inject planetesimals at a lower rate compared to the compact disc formation timescale, then these systems would be capable of hosting compact discs. Such compact discs would persist until dispersed by some future tidal disruption event of comparable progenitor mass or dispersed by intrinsic processes on some characteristic timescale.
	
	Another possibility is that occasionally massive planetesimals are injected to form unusually massive compact discs compared with the typical planetesimal masses. Such a compact disc would resist a catastrophic fate until the next tidal disruption of a similarly massive planetesimal. As long as it encounters injected planetesimals of significantly less mass, it would be able to fully circularize the ensuing tidal streams. How long such a compact disc may persist depends on the balance between its intrinsic lifetime (rate of mass loss to WD accretion) and the mass contribution from fragment self-erosion (i.e the tidal streams also contribute their own mass, as we discuss next).
	
	The possibilities above lack the details to support a decisive conclusion among the two possibilities, however the point we wish to emphasize is that the scarce observations of dusty compact discs in combination with our model, generally imply a dependence on either planetesimal injection rate or size distribution. 
	
	We also suggested that in the absence of a compact disc, the accretion rate onto the WD might be steadily controlled by PR drag. However, in the presence of a compact disc interacting with a tidal stream, the rate of accretion might increase via brief and intense accretion episodes from tidal disruption events. Similar ideas were suggested by \cite{Jura-2008,FarihiEtAl-2012} in order to explain why polluted He-dominated WDs exhibit average accretion rates that exceed that of polluted H-dominated WDs by one or two orders of magnitudes, while the rate in H-dominated WDs is considered to be in agreement with the theoretical mass transfer limit via PR drag \citep{GirvenEtAl-2012}. We note that other ideas were also suggested, such as a runway accretion model \citep{Rafikov-2011} and differential volatile accretion in rare water-bearing planetesimals \citep{MalamudPerets-2016}.

	\subsection{Erosion vs. circularization timescales}\label{SS:ErosionVsCircularization}
	In section \ref{SS:GasProduction} we identify two fragment erosion channels. In a gaseous compact disc we have sputtering, producing a yield $Y$ in the range 0.1-1. In a dusty compact disc we have shattering, producing an effective yield of $\sim$1. By simple arguments, we may now estimate if fragment self-erosion can potentially hinder the circularization process.
	
	For a fragment of mass $m$, radius $R$ and density $\rho$ we have $m=4\pi\rho R^3/3$. The yield is by definition the eroded fragment mass over the incident compact disc mass, and therefore if it crosses the compact disc face-on, $\Delta m$ is $\pi R^2 \Sigma Y$. We now have
	
	\begin{equation}\label{eq:dRdt}
		\frac{dR}{dt}=\frac{\Delta m}{\Delta t} \frac{dR}{dm}=\frac{\Sigma Y}{8 \pi \rho} \sqrt{\frac{GM_{\rm WD}}{a^3}},
	\end{equation}
	\noindent where $\Delta t=P=2\pi \sqrt{a^3/GM_{\rm WD}}$ is the orbital period. Using Equation \ref{eq:dRdt} we may obtain the erosion timescale $\tau_{\rm erosion}$ from
	
	\begin{equation}\label{eq:erosionTimescale}
		\tau_{\rm erosion}=\frac{R}{dR/dt}=\frac{8 \pi \rho R}{\Sigma Y} \sqrt{\frac{a^3}{GM_{\rm WD}}}.
	\end{equation}

	Erosion outpaces circularization when, using Equation \ref{eq:tcirc}, 
	
	\begin{equation}\label{eq:erosionTimescale}
		\frac{\tau_{\rm erosion}}{\tau_{\rm circ}}<1.
	\end{equation}
	 
	Or equivalently, when
	
	\begin{equation}\label{eq:erosionTimescale}
		Y>6\frac{a}{q}.
	\end{equation}

	Even if we take the minimum semi-major axis $a=0.1$ AU, and the maximum pericentre $q=1 R_{\odot}$, we require $Y>129$ for the erosion timescale to replace the circularization timescale. Since we have found that $Y\le 1$, we conclude that the circularization time is essentially unaffected by erosion.
	
	We also conclude that for partial circularization, erosion is entirely negligible. We have shown in Section \ref{SS:PartialCircularization} that even when $M_{\rm progenitor}$ is many orders of magnitude larger than the compact disc mass $M$, the tidal stream undergoes significant circularization. The mass eroded away (collectively) from the tidal stream is roughly equal to $MY$, hence it would be negligible compared to the remaining mass in the stream.
	
	Finally, if instead $M \gg M_{\rm progenitor}$ and full circularization ensues, Equation \ref{eq:FullCircularization} shows that the tidal stream encounters nearly its own mass in the compact disc. If the yield $Y$ is equal or close to 1, as we have found, by definition the tidal stream is greatly eroded by the time it circularizes since the eroded mass is the mass encountered in the compact disc times $\sim$1. Fully circularized fragments embedded in the disc will always turn out to be a small fraction of their initial mass by the time they circularize. Additionally, eroded gas and dust grains quickly conjoin the compact disc (as in Section \ref{SS:ObservationalConsequences}).

	\section{Summary}\label{S:summary}
	Various theoretical studies have shown that planetary tidal disruptions by WDs initially form extremely eccentric fragment streams, with $e>0.9$ and often approaching unity. Around a subset of systems, observational evidence of infrared excesses indicates that some structures eventually shrink to form compact, nearly circular discs. Further observational support comes from rare compact discs that have a gaseous component. Explaining why this change occurs poses a major theoretical challenge, as the tidal streams need to lose large amounts of both energy and angular momentum. Previous theoretical works have suggested the reprocessing of stellar light, which can alter the orbits of the tidal stream constituents either by PR drag or the Yarkovsky force. However the former requires that the stream is primarily composed of small dust, which is (at least initially) incorrect, and the latter remains to be thoroughly explored. In this paper, we suggest a new mechanism: drag-assisted circularization as an effective alternative in systems that posses a circumstellar disc.
	
	Our model is based around the observation that these dust/gas discs around WDs are typically compact and located within the stellar Roche radius. This means that fragments from subsequent tidal disruptions will collide with this material at their pericenter, leading to high velocity collisions that generate strong drag forces on the fragments, while carving gaps in the compact disc. We calculate the resulting changes in the fragment's orbits with an analytical model, where we account for the diffusive filling of the gaps. We then verify our expressions with more detailed N-body simulations.
	
	Our results indicate that drag-assisted circularization is a rapid process that can completely circularize large fragments up to a kilometre in size within the expected disc lifetimes. This efficiency is primarily due to the high pericentre velocities at near-unity ecccentricity, that greatly speed up the process. We find that full circularization of tidal fragments can occur if the compact disc mass is as low as $10^{16}$ g. Interestingly, drag-assisted circularization is efficient for fragments on a wide range of semi-major axes, as the longer orbital timescales at wider orbits are compensated by their increased perictenter velocity. 
	
	We also find that the outcome of drag-based circularization is primarily determined by the mass ratio of the compact disc and stream progenitor. This comes from the fact that complete circuarization requires a compact disc that is ideally much more massive than the tidal stream, in which case the tidal stream would eventually conjoin the compact disc. Significant partial circularization is on the other hand possible even when the tidal stream exceeds the compact disc by up to seven orders of magnitude in mass, but then the compact disc is dispersed, and continued circularization requires other mechanisms. We therefore expect partial drag-assisted circularization to be a ubiquitous feature of the fragment's orbital evolution, as even small progenitors that form the pre-existing compact disc can significantly affect the orbits of major bodies. We speculate that because the scale of partial circularization is also a function of fragment size, it can trigger a subsequent phase of collisions between fragments that grind down the remaining material, facilitating further circularization by PR drag. We will investigate this further in a future work.
	
	Besides drag, the high collision velocities between fragments and the compact disc necessarily lead to the production of significant amounts of gas mainly through vaporization of pre-existing compact disc dust and dust eroded from the fragments, but also via other channels. Given a size-distribution of tidal stream fragments, gas production starts off continuously, yielding an observable signature of ongoing early circularization. But as the smaller fragments circularize their collision velocities decline, the gas production then originates from increasingly larger fragments and thus it becomes increasingly intermittent and harder to observe. Gas production finally halts completely when either all the tidal fragments are circularized or the compact disc has dissipated. In this way, our model offers a natural explanation for the fact that gaseous material is only observed around a small fraction of polluted WDs.
	
	We also speculate that fragments can scatter material far from the WD, giving rise to a dusty halo which evolves towards the WD through PR drag until eventually the material would be re-assimilated inside the compact disc. The timescale could be approximately compatible with recent observational evidence for infrared variability.
	
	A necessary requirement of the model is the pre-existence of a compact disc around the WD. We show that this problem is not insurmountable since a number of mechanisms may enable its formation in the first place.

	\section*{Acknowledgements}
	We would like to thank Jay Farihi and Na'ama Hallakoun for extremely helpful discussions relating to the observational constraints associated with debris discs, and also Christopher Manser for helpful references regarding observed systems. We would like to thank anonymous referee for highly useful suggestions that greatly improved the quality of the paper. UM acknowledges support from the Pazy Foundation grant 01023882.
	
	\section*{Data availability}
	The simulation data that support the findings of this study are available upon request from the corresponding author, UM.
	
	
	
	\newpage
	\bibliographystyle{mnras}
	\bibliography{bibfile} 

\begin{thebibliography}{}
\makeatletter
\relax
\def\mn@urlcharsother{\let\do\@makeother \do\$\do\&\do\#\do\^\do\_\do\%\do\~}
\def\mn@doi{\begingroup\mn@urlcharsother \@ifnextchar [ {\mn@doi@}
  {\mn@doi@[]}}
\def\mn@doi@[#1]#2{\def\@tempa{#1}\ifx\@tempa\@empty \href
  {http://dx.doi.org/#2} {doi:#2}\else \href {http://dx.doi.org/#2} {#1}\fi
  \endgroup}
\def\mn@eprint#1#2{\mn@eprint@#1:#2::\@nil}
\def\mn@eprint@arXiv#1{\href {http://arxiv.org/abs/#1} {{\tt arXiv:#1}}}
\def\mn@eprint@dblp#1{\href {http://dblp.uni-trier.de/rec/bibtex/#1.xml}
  {dblp:#1}}
\def\mn@eprint@#1:#2:#3:#4\@nil{\def\@tempa {#1}\def\@tempb {#2}\def\@tempc
  {#3}\ifx \@tempc \@empty \let \@tempc \@tempb \let \@tempb \@tempa \fi \ifx
  \@tempb \@empty \def\@tempb {arXiv}\fi \@ifundefined
  {mn@eprint@\@tempb}{\@tempb:\@tempc}{\expandafter \expandafter \csname
  mn@eprint@\@tempb\endcsname \expandafter{\@tempc}}}

\bibitem[\protect\citeauthoryear{{Ahrens} \& {O'Keefe}}{{Ahrens} \&
  {O'Keefe}}{1972}]{AhrensOkeefe-1972}
{Ahrens} T.~J.,  {O'Keefe} J.~D.,  1972, \mn@doi [Moon] {10.1007/BF00562927},
  \href {https://ui.adsabs.harvard.edu/abs/1972Moon....4..214A} {4, 214}

\bibitem[\protect\citeauthoryear{{Bear} \& {Soker}}{{Bear} \&
  {Soker}}{2015}]{BearSoker-2015}
{Bear} E.,  {Soker} N.,  2015, \mn@doi [Monthly Notices of the Royal
  Astronomical Society] {10.1093/mnras/stv921}, \href
  {https://ui.adsabs.harvard.edu/abs/2015MNRAS.450.4233B} {450, 4233}

\bibitem[\protect\citeauthoryear{{Behrisch} \& {Eckstein}}{{Behrisch} \&
  {Eckstein}}{2007}]{BehrischEckstein-2007}
{Behrisch} R.,  {Eckstein} W.,  2007, {Sputtering by Particle Bombardment}.
Springer

\bibitem[\protect\citeauthoryear{{Biele} et~al.,}{{Biele}
  et~al.}{2015}]{BieleEtAl-2015}
{Biele} J.,  et~al., 2015, \mn@doi [Science] {10.1126/science.aaa9816}, \href
  {https://ui.adsabs.harvard.edu/abs/2015Sci...349a9816B} {349, 1.9816}

\bibitem[\protect\citeauthoryear{{Bochkarev} \& {Rafikov}}{{Bochkarev} \&
  {Rafikov}}{2011}]{BochkarevRafikov-2011}
{Bochkarev} K.~V.,  {Rafikov} R.~R.,  2011, \mn@doi [The Astrophysical Journal]
  {10.1088/0004-637X/741/1/36}, \href
  {https://ui.adsabs.harvard.edu/abs/2011ApJ...741...36B} {741, 36}

\bibitem[\protect\citeauthoryear{{Bonsor}, {Mustill}  \& {Wyatt}}{{Bonsor}
  et~al.}{2011}]{BonsorEtAl-2011}
{Bonsor} A.,  {Mustill} A.~J.,   {Wyatt} M.~C.,  2011, \mn@doi [Monthly Notices
  of the Royal Astronomical Society] {10.1111/j.1365-2966.2011.18524.x}, \href
  {http://adsabs.harvard.edu/abs/2011MNRAS.414..930B} {414, 930}

\bibitem[\protect\citeauthoryear{{Bonsor}, {Farihi}, {Wyatt}  \& {van
  Lieshout}}{{Bonsor} et~al.}{2017}]{BonsorEtAl-2017}
{Bonsor} A.,  {Farihi} J.,  {Wyatt} M.~C.,   {van Lieshout} R.,  2017, \mn@doi
  [Monthly Notices of the Royal Astronomical Society] {10.1093/mnras/stx425},
  \href {https://ui.adsabs.harvard.edu/abs/2017MNRAS.468..154B} {468, 154}

\bibitem[\protect\citeauthoryear{{Brinkworth}, {G{\"a}nsicke}, {Marsh}, {Hoard}
   \& {Tappert}}{{Brinkworth} et~al.}{2009}]{BrinkworthEtAl-2009}
{Brinkworth} C.~S.,  {G{\"a}nsicke} B.~T.,  {Marsh} T.~R.,  {Hoard} D.~W.,
  {Tappert} C.,  2009, \mn@doi [The Astrophysical Journal]
  {10.1088/0004-637X/696/2/1402}, \href
  {https://ui.adsabs.harvard.edu/abs/2009ApJ...696.1402B} {696, 1402}

\bibitem[\protect\citeauthoryear{{Bromley} \& {Kenyon}}{{Bromley} \&
  {Kenyon}}{2013}]{BromleyKenyon-2013}
{Bromley} B.~C.,  {Kenyon} S.~J.,  2013, \mn@doi [The Astrophysical Journal]
  {10.1088/0004-637X/764/2/192}, \href
  {https://ui.adsabs.harvard.edu/abs/2013ApJ...764..192B} {764, 192}

\bibitem[\protect\citeauthoryear{{Brown}, {Veras}  \& {G{\"a}nsicke}}{{Brown}
  et~al.}{2017}]{BrownEtAl-2017}
{Brown} J.~C.,  {Veras} D.,   {G{\"a}nsicke} B.~T.,  2017, \mn@doi [Monthly
  Notices of the Royal Astronomical Society] {10.1093/mnras/stx428}, \href
  {http://adsabs.harvard.edu/abs/2017MNRAS.468.1575B} {468, 1575}

\bibitem[\protect\citeauthoryear{{Caiazzo} \& {Heyl}}{{Caiazzo} \&
  {Heyl}}{2017}]{CaiazzoHeyl-2017}
{Caiazzo} I.,  {Heyl} J.~S.,  2017, \mn@doi [Monthly Notices of the Royal
  Astronomical Society] {10.1093/mnras/stx1036}, \href
  {http://adsabs.harvard.edu/abs/2017MNRAS.469.2750C} {469, 2750}

\bibitem[\protect\citeauthoryear{{Cauley}, {Farihi}, {Redfield}, {Bachman},
  {Parsons}  \& {G{\"a}nsicke}}{{Cauley} et~al.}{2018}]{CauleyEtAl-2018}
{Cauley} P.~W.,  {Farihi} J.,  {Redfield} S.,  {Bachman} S.,  {Parsons} S.~G.,
   {G{\"a}nsicke} B.~T.,  2018, \mn@doi [The Astrophysical Journal Letters]
  {10.3847/2041-8213/aaa3d9}, \href
  {https://ui.adsabs.harvard.edu/abs/2018ApJ...852L..22C} {852, L22}

\bibitem[\protect\citeauthoryear{{D'Angelo} \& {Podolak}}{{D'Angelo} \&
  {Podolak}}{2015}]{Podolak2015}
{D'Angelo} G.,  {Podolak} M.,  2015, \mn@doi [The Astrophysical Journal]
  {10.1088/0004-637X/806/2/203}, \href
  {http://adsabs.harvard.edu/abs/2015ApJ...806..203D} {806, 203}

\bibitem[\protect\citeauthoryear{{Debes} \& {Sigurdsson}}{{Debes} \&
  {Sigurdsson}}{2002}]{DebesSigurdsson-2002}
{Debes} J.~H.,  {Sigurdsson} S.,  2002, \mn@doi [The Astrophysical Journal]
  {10.1086/340291}, \href {http://adsabs.harvard.edu/abs/2002ApJ...572..556D}
  {572, 556}

\bibitem[\protect\citeauthoryear{{Debes}, {Walsh}  \& {Stark}}{{Debes}
  et~al.}{2012}]{DebesEtAl-2012}
{Debes} J.~H.,  {Walsh} K.~J.,   {Stark} C.,  2012, \mn@doi [The Astrophysical
  Journal] {10.1088/0004-637X/747/2/148}, \href
  {http://adsabs.harvard.edu/abs/2012ApJ...747..148D} {747, 148}

\bibitem[\protect\citeauthoryear{{Dennihy}, {Debes}, {Dunlap}, {Dufour},
  {Teske}  \& {Clemens}}{{Dennihy} et~al.}{2016}]{DennihyEtAl-2016}
{Dennihy} E.,  {Debes} J.~H.,  {Dunlap} B.~H.,  {Dufour} P.,  {Teske} J.~K.,
  {Clemens} J.~C.,  2016, \mn@doi [The Astrophysical Journal]
  {10.3847/0004-637X/831/1/31}, \href
  {https://ui.adsabs.harvard.edu/abs/2016ApJ...831...31D} {831, 31}

\bibitem[\protect\citeauthoryear{{Dennihy}, {Clemens}, {Dunlap}, {Fanale},
  {Fuchs}  \& {Hermes}}{{Dennihy} et~al.}{2018}]{DennihyEtAl-2018}
{Dennihy} E.,  {Clemens} J.~C.,  {Dunlap} B.~H.,  {Fanale} S.~M.,  {Fuchs}
  J.~T.,   {Hermes} J.~J.,  2018, \mn@doi [The Astrophysical Journal]
  {10.3847/1538-4357/aaa89b}, \href
  {https://ui.adsabs.harvard.edu/abs/2018ApJ...854...40D} {854, 40}

\bibitem[\protect\citeauthoryear{{Dennihy} et~al.,}{{Dennihy}
  et~al.}{2020}]{DennihyEtAl-2020}
{Dennihy} E.,  et~al., 2020, arXiv e-prints, \href
  {https://ui.adsabs.harvard.edu/abs/2020arXiv201003693D} {p. arXiv:2010.03693}

\bibitem[\protect\citeauthoryear{{Farihi}}{{Farihi}}{2016}]{Farihi-2016}
{Farihi} J.,  2016, \mn@doi [New Astronomy Reviews]
  {10.1016/j.newar.2016.03.001}, \href
  {http://adsabs.harvard.edu/abs/2016NewAR..71....9F} {71, 9}

\bibitem[\protect\citeauthoryear{{Farihi}, {Jura}  \& {Zuckerman}}{{Farihi}
  et~al.}{2009}]{FarihiEtAl-2009}
{Farihi} J.,  {Jura} M.,   {Zuckerman} B.,  2009, \mn@doi [The Astrophysical
  Journal] {10.1088/0004-637X/694/2/805}, \href
  {https://ui.adsabs.harvard.edu/abs/2009ApJ...694..805F} {694, 805}

\bibitem[\protect\citeauthoryear{{Farihi}, {Jura}, {Lee}  \&
  {Zuckerman}}{{Farihi} et~al.}{2010}]{FarihiEtAl-2010}
{Farihi} J.,  {Jura} M.,  {Lee} J.~E.,   {Zuckerman} B.,  2010, \mn@doi [The
  Astrophysical Journal] {10.1088/0004-637X/714/2/1386}, \href
  {https://ui.adsabs.harvard.edu/abs/2010ApJ...714.1386F} {714, 1386}

\bibitem[\protect\citeauthoryear{{Farihi}, {G{\"a}nsicke}, {Wyatt}, {Girven},
  {Pringle}  \& {King}}{{Farihi} et~al.}{2012}]{FarihiEtAl-2012}
{Farihi} J.,  {G{\"a}nsicke} B.~T.,  {Wyatt} M.~C.,  {Girven} J.,  {Pringle}
  J.~E.,   {King} A.~R.,  2012, \mn@doi [Monthly Notices of the Royal
  Astronomical Society] {10.1111/j.1365-2966.2012.21215.x}, \href
  {http://adsabs.harvard.edu/abs/2012MNRAS.424..464F} {424, 464}

\bibitem[\protect\citeauthoryear{{Farihi} et~al.,}{{Farihi}
  et~al.}{2018}]{FarihiEtAl-2018}
{Farihi} J.,  et~al., 2018, \mn@doi [Monthly Notices of the Royal Astronomical
  Society] {10.1093/mnras/sty2331}, \href
  {https://ui.adsabs.harvard.edu/abs/2018MNRAS.481.2601F} {481, 2601}

\bibitem[\protect\citeauthoryear{{Fortin-Archambault}, {Dufour}  \&
  {Xu}}{{Fortin-Archambault} et~al.}{2020}]{FortinArchambaultEtAl-2020}
{Fortin-Archambault} M.,  {Dufour} P.,   {Xu} S.,  2020, \mn@doi [The
  Astrophysical Journal] {10.3847/1538-4357/ab585a}, \href
  {https://ui.adsabs.harvard.edu/abs/2020ApJ...888...47F} {888, 47}

\bibitem[\protect\citeauthoryear{{G{\"a}nsicke}, {Marsh}, {Southworth}  \&
  {Rebassa-Mansergas}}{{G{\"a}nsicke} et~al.}{2006}]{GansickeEtAl-2006}
{G{\"a}nsicke} B.~T.,  {Marsh} T.~R.,  {Southworth} J.,   {Rebassa-Mansergas}
  A.,  2006, \mn@doi [Science] {10.1126/science.1135033}, \href
  {http://adsabs.harvard.edu/abs/2006Sci...314.1908G} {314, 1908}

\bibitem[\protect\citeauthoryear{{G{\"a}nsicke}, {Koester}, {Marsh},
  {Rebassa-Mansergas}  \& {Southworth}}{{G{\"a}nsicke}
  et~al.}{2008}]{GansickeEtAl-2008}
{G{\"a}nsicke} B.~T.,  {Koester} D.,  {Marsh} T.~R.,  {Rebassa-Mansergas} A.,
  {Southworth} J.,  2008, \mn@doi [Monthly Notices of the Royal Astronomical
  Society] {10.1111/j.1745-3933.2008.00565.x}, \href
  {http://adsabs.harvard.edu/abs/2008MNRAS.391L.103G} {391, L103}

\bibitem[\protect\citeauthoryear{{G{\"a}nsicke}, {Koester}, {Farihi}, {Girven},
  {Parsons}  \& {Breedt}}{{G{\"a}nsicke} et~al.}{2012}]{GansickeEtAl-2012}
{G{\"a}nsicke} B.~T.,  {Koester} D.,  {Farihi} J.,  {Girven} J.,  {Parsons}
  S.~G.,   {Breedt} E.,  2012, \mn@doi [Monthly Notices of the Royal
  Astronomical Society] {10.1111/j.1365-2966.2012.21201.x}, \href
  {https://ui.adsabs.harvard.edu/abs/2012MNRAS.424..333G} {424, 333}

\bibitem[\protect\citeauthoryear{{Gentile Fusillo} et~al.,}{{Gentile Fusillo}
  et~al.}{2020}]{Gentile-FusilloEtAl-2020}
{Gentile Fusillo} N.~P.,  et~al., 2020, arXiv e-prints, \href
  {https://ui.adsabs.harvard.edu/abs/2020arXiv201013807G} {p. arXiv:2010.13807}

\bibitem[\protect\citeauthoryear{{Girven}, {Brinkworth}, {Farihi},
  {G{\"a}nsicke}, {Hoard}, {Marsh}  \& {Koester}}{{Girven}
  et~al.}{2012}]{GirvenEtAl-2012}
{Girven} J.,  {Brinkworth} C.~S.,  {Farihi} J.,  {G{\"a}nsicke} B.~T.,  {Hoard}
  D.~W.,  {Marsh} T.~R.,   {Koester} D.,  2012, \mn@doi [The Astrophysical
  Journal] {10.1088/0004-637X/749/2/154}, \href
  {http://adsabs.harvard.edu/abs/2012ApJ...749..154G} {749, 154}

\bibitem[\protect\citeauthoryear{{Goldreich} \& {Soter}}{{Goldreich} \&
  {Soter}}{1966}]{GoldreichSoter-1966}
{Goldreich} P.,  {Soter} S.,  1966, \mn@doi [Icarus]
  {10.1016/0019-1035(66)90051-0}, \href
  {https://ui.adsabs.harvard.edu/abs/1966Icar....5..375G} {5, 375}

\bibitem[\protect\citeauthoryear{{Goldreich} \& {Tremaine}}{{Goldreich} \&
  {Tremaine}}{1978}]{GoldreichTremaine-1978}
{Goldreich} P.,  {Tremaine} S.~D.,  1978, \mn@doi [Icarus]
  {10.1016/0019-1035(78)90164-1}, \href
  {https://ui.adsabs.harvard.edu/abs/1978Icar...34..227G} {34, 227}

\bibitem[\protect\citeauthoryear{{Graham}, {Matthews}, {Neugebauer}  \&
  {Soifer}}{{Graham} et~al.}{1990}]{GrahamEtAl-1990}
{Graham} J.~R.,  {Matthews} K.,  {Neugebauer} G.,   {Soifer} B.~T.,  1990,
  \mn@doi [The Astrophysical Journal] {10.1086/168907}, \href
  {https://ui.adsabs.harvard.edu/abs/1990ApJ...357..216G} {357, 216}

\bibitem[\protect\citeauthoryear{{Grishin} \& {Perets}}{{Grishin} \&
  {Perets}}{2015}]{GrishinPerets2015}
{Grishin} E.,  {Perets} H.~B.,  2015, \mn@doi [The Astrophysical Journal]
  {10.1088/0004-637X/811/1/54}, \href
  {https://ui.adsabs.harvard.edu/abs/2015ApJ...811...54G} {811, 54}

\bibitem[\protect\citeauthoryear{{Grishin} \& {Veras}}{{Grishin} \&
  {Veras}}{2019}]{GrishinVeras-2019}
{Grishin} E.,  {Veras} D.,  2019, \mn@doi [Monthly Notices of the Royal
  Astronomical Society] {10.1093/mnras/stz2148}, \href
  {https://ui.adsabs.harvard.edu/abs/2019MNRAS.489..168G} {489, 168}

\bibitem[\protect\citeauthoryear{{Grishin}, {Perets}  \& {Avni}}{{Grishin}
  et~al.}{2019}]{GrishinEtAl-2019}
{Grishin} E.,  {Perets} H.~B.,   {Avni} Y.,  2019, \mn@doi [Monthly Notices of
  the Royal Astronomical Society] {10.1093/mnras/stz1505}, \href
  {https://ui.adsabs.harvard.edu/abs/2019MNRAS.487.3324G} {487, 3324}

\bibitem[\protect\citeauthoryear{{Grishin}, {Malamud}, {Perets}, {Wand el}  \&
  {Sch{\"a}fer}}{{Grishin} et~al.}{2020a}]{GrishinEtAl-2020a}
{Grishin} E.,  {Malamud} U.,  {Perets} H.~B.,  {Wand el} O.,   {Sch{\"a}fer}
  C.~M.,  2020a, \mn@doi [Nature] {10.1038/s41586-020-2194-z}, \href
  {https://ui.adsabs.harvard.edu/abs/2020Natur.580..463G} {580, 463}

\bibitem[\protect\citeauthoryear{{Grishin}, {Rozner}  \& {Perets}}{{Grishin}
  et~al.}{2020b}]{GrishinEtAl-2020b}
{Grishin} E.,  {Rozner} M.,   {Perets} H.~B.,  2020b, \mn@doi [The
  Astrophysical Journal Letters] {10.3847/2041-8213/aba266}, \href
  {https://ui.adsabs.harvard.edu/abs/2020ApJ...898L..13G} {898, L13}

\bibitem[\protect\citeauthoryear{{Hamers} \& {Portegies Zwart}}{{Hamers} \&
  {Portegies Zwart}}{2016}]{HamersPortegiesZwart-2016}
{Hamers} A.~S.,  {Portegies Zwart} S.~F.,  2016, \mn@doi [Monthly Notices of
  the Royal Astronomical Society] {10.1093/mnrasl/slw134}, \href
  {http://adsabs.harvard.edu/abs/2016MNRAS.462L..84H} {462, L84}

\bibitem[\protect\citeauthoryear{{Hameury}}{{Hameury}}{2020}]{Hameury-2020}
{Hameury} J.~M.,  2020, \mn@doi [Advances in Space Research]
  {10.1016/j.asr.2019.10.022}, \href
  {https://ui.adsabs.harvard.edu/abs/2020AdSpR..66.1004H} {66, 1004}

\bibitem[\protect\citeauthoryear{{Hut}}{{Hut}}{1981}]{Hut-1981}
{Hut} P.,  1981, Astronomy \& Astrophysics, \href
  {https://ui.adsabs.harvard.edu/abs/1981A&A....99..126H} {99, 126}

\bibitem[\protect\citeauthoryear{{Izquierdo} et~al.,}{{Izquierdo}
  et~al.}{2018}]{IzquierdoEtAl-2018}
{Izquierdo} P.,  et~al., 2018, \mn@doi [Monthly Notices of the Royal
  Astronomical Society] {10.1093/mnras/sty2315}, \href
  {https://ui.adsabs.harvard.edu/abs/2018MNRAS.481..703I} {481, 703}

\bibitem[\protect\citeauthoryear{{Jura}}{{Jura}}{2003}]{Jura-2003}
{Jura} M.,  2003, \mn@doi [The Astrophysical Journal] {10.1086/374036}, \href
  {http://adsabs.harvard.edu/abs/2003ApJ...584L..91J} {584, L91}

\bibitem[\protect\citeauthoryear{{Jura}}{{Jura}}{2008}]{Jura-2008}
{Jura} M.,  2008, \mn@doi [The Astronomical Journal]
  {10.1088/0004-6256/135/5/1785}, \href
  {http://adsabs.harvard.edu/abs/2008AJ....135.1785J} {135, 1785}

\bibitem[\protect\citeauthoryear{{Karjalainen}, {de Mooij}, {Karjalainen}  \&
  {Gibson}}{{Karjalainen} et~al.}{2019}]{KarjalainenEtAl-2019}
{Karjalainen} M.,  {de Mooij} E. J.~W.,  {Karjalainen} R.,   {Gibson} N.~P.,
  2019, \mn@doi [Monthly Notices of the Royal Astronomical Society]
  {10.1093/mnras/sty2778}, \href
  {https://ui.adsabs.harvard.edu/abs/2019MNRAS.482..999K} {482, 999}

\bibitem[\protect\citeauthoryear{{Kenyon} \& {Bromley}}{{Kenyon} \&
  {Bromley}}{2017a}]{KenyonBromley-2017a}
{Kenyon} S.~J.,  {Bromley} B.~C.,  2017a, \mn@doi [The Astrophysical Journal]
  {10.3847/1538-4357/aa7b85}, \href
  {https://ui.adsabs.harvard.edu/abs/2017ApJ...844..116K} {844, 116}

\bibitem[\protect\citeauthoryear{{Kenyon} \& {Bromley}}{{Kenyon} \&
  {Bromley}}{2017b}]{KenyonBromley-2017b}
{Kenyon} S.~J.,  {Bromley} B.~C.,  2017b, \mn@doi [The Astrophysical Journal]
  {10.3847/1538-4357/aa9570}, \href
  {https://ui.adsabs.harvard.edu/abs/2017ApJ...850...50K} {850, 50}

\bibitem[\protect\citeauthoryear{{Kilic}, {von Hippel}, {Leggett}  \&
  {Winget}}{{Kilic} et~al.}{2006}]{KilicEtAl-2006}
{Kilic} M.,  {von Hippel} T.,  {Leggett} S.~K.,   {Winget} D.~E.,  2006,
  \mn@doi [The Astrophysical Journal] {10.1086/504682}, \href
  {http://adsabs.harvard.edu/abs/2006ApJ...646..474K} {646, 474}

\bibitem[\protect\citeauthoryear{{Koester}}{{Koester}}{2009}]{Koester-2009}
{Koester} D.,  2009, \mn@doi [Astronomy and Astrophysics]
  {10.1051/0004-6361/200811468}, \href
  {http://adsabs.harvard.edu/abs/2009A%26A...498..517K} {498, 517}

\bibitem[\protect\citeauthoryear{{Koester}, {G{\"a}nsicke}  \&
  {Farihi}}{{Koester} et~al.}{2014}]{KoesterEtAl-2014}
{Koester} D.,  {G{\"a}nsicke} B.~T.,   {Farihi} J.,  2014, \mn@doi [Astronomy
  and Astrophysics] {10.1051/0004-6361/201423691}, \href
  {http://adsabs.harvard.edu/abs/2014A%26A...566A..34K} {566, A34}

\bibitem[\protect\citeauthoryear{{Koschny} \& {Gr{\"u}n}}{{Koschny} \&
  {Gr{\"u}n}}{2001}]{KoschnyGrun-2001}
{Koschny} D.,  {Gr{\"u}n} E.,  2001, \mn@doi [Icarus] {10.1006/icar.2001.6708},
  \href {https://ui.adsabs.harvard.edu/abs/2001Icar..154..402K} {154, 402}

\bibitem[\protect\citeauthoryear{{Kratter} \& {Perets}}{{Kratter} \&
  {Perets}}{2012}]{KratterPerets-2012}
{Kratter} K.~M.,  {Perets} H.~B.,  2012, \mn@doi [The Astrophysical Journal]
  {10.1088/0004-637X/753/1/91}, \href
  {http://adsabs.harvard.edu/abs/2012ApJ...753...91K} {753, 91}

\bibitem[\protect\citeauthoryear{Kudriavtsev, Villegas, Godines  \&
  Asomoza}{Kudriavtsev et~al.}{2005}]{KudriavtsevEtAl-2005}
Kudriavtsev Y.,  Villegas A.,  Godines A.,   Asomoza R.,  2005, \mn@doi
  [Applied Surface Science] {https://doi.org/10.1016/j.apsusc.2004.06.014},
  239, 273

\bibitem[\protect\citeauthoryear{{Leinhardt} \& {Stewart}}{{Leinhardt} \&
  {Stewart}}{2009}]{LeinhardtStewart-2009}
{Leinhardt} Z.~M.,  {Stewart} S.~T.,  2009, \mn@doi [Icarus]
  {10.1016/j.icarus.2008.09.013}, \href
  {https://ui.adsabs.harvard.edu/abs/2009Icar..199..542L} {199, 542}

\bibitem[\protect\citeauthoryear{{Leitch-Devlin} \& {Williams}}{{Leitch-Devlin}
  \& {Williams}}{1985}]{Leitch-DevlinWilliams-1985}
{Leitch-Devlin} M.~A.,  {Williams} D.~A.,  1985, \mn@doi [Monthly Notices of
  the Royal Astronomical Society] {10.1093/mnras/213.2.295}, \href
  {https://ui.adsabs.harvard.edu/abs/1985MNRAS.213..295L} {213, 295}

\bibitem[\protect\citeauthoryear{{Malamud} \& {Perets}}{{Malamud} \&
  {Perets}}{2016}]{MalamudPerets-2016}
{Malamud} U.,  {Perets} H.~B.,  2016, \mn@doi [The Astrophysical Journal]
  {10.3847/0004-637X/832/2/160}, \href
  {http://adsabs.harvard.edu/abs/2016ApJ...832..160M} {832, 160}

\bibitem[\protect\citeauthoryear{{Malamud} \& {Perets}}{{Malamud} \&
  {Perets}}{2017a}]{MalamudPerets-2017a}
{Malamud} U.,  {Perets} H.~B.,  2017a, \mn@doi [The Astrophysical Journal]
  {10.3847/1538-4357/aa7055}, \href
  {http://adsabs.harvard.edu/abs/2017ApJ...842...67M} {842, 67}

\bibitem[\protect\citeauthoryear{{Malamud} \& {Perets}}{{Malamud} \&
  {Perets}}{2017b}]{MalamudPerets-2017b}
{Malamud} U.,  {Perets} H.~B.,  2017b, \mn@doi [The Astrophysical Journal]
  {10.3847/1538-4357/aa8df5}, \href
  {http://adsabs.harvard.edu/abs/2017ApJ...849....8M} {849, 8}

\bibitem[\protect\citeauthoryear{{Malamud} \& {Perets}}{{Malamud} \&
  {Perets}}{2020a}]{MalamudPerets-2020a}
{Malamud} U.,  {Perets} H.~B.,  2020a, \mn@doi [Monthly Notices of the Royal
  Astronomical Society] {10.1093/mnras/staa142}, \href
  {https://ui.adsabs.harvard.edu/abs/2020MNRAS.492.5561M} {492, 5561}

\bibitem[\protect\citeauthoryear{{Malamud} \& {Perets}}{{Malamud} \&
  {Perets}}{2020b}]{MalamudPerets-2020b}
{Malamud} U.,  {Perets} H.~B.,  2020b, \mn@doi [Monthly Notices of the Royal
  Astronomical Society] {10.1093/mnras/staa143}, \href
  {https://ui.adsabs.harvard.edu/abs/2020MNRAS.493..698M} {493, 698}

\bibitem[\protect\citeauthoryear{{Manser} et~al.,}{{Manser}
  et~al.}{2016a}]{ManserEtAl-2016a}
{Manser} C.~J.,  et~al., 2016a, \mn@doi [Monthly Notices of the Royal
  Astronomical Society] {10.1093/mnras/stv2603}, \href
  {http://adsabs.harvard.edu/abs/2016MNRAS.455.4467M} {455, 4467}

\bibitem[\protect\citeauthoryear{{Manser}, {G{\"a}nsicke}, {Koester}, {Marsh}
  \& {Southworth}}{{Manser} et~al.}{2016b}]{ManserEtAl-2016b}
{Manser} C.~J.,  {G{\"a}nsicke} B.~T.,  {Koester} D.,  {Marsh} T.~R.,
  {Southworth} J.,  2016b, \mn@doi [Monthly Notices of the Royal Astronomical
  Society] {10.1093/mnras/stw1760}, \href
  {https://ui.adsabs.harvard.edu/abs/2016MNRAS.462.1461M} {462, 1461}

\bibitem[\protect\citeauthoryear{{Manser} et~al.,}{{Manser}
  et~al.}{2019}]{ManserEtAl-2019}
{Manser} C.~J.,  et~al., 2019, \mn@doi [Science] {10.1126/science.aat5330},
  \href {https://ui.adsabs.harvard.edu/abs/2019Sci...364...66M} {364, 66}

\bibitem[\protect\citeauthoryear{{Manser}, {G{\"a}nsicke}, {Gentile Fusillo},
  {Ashley}, {Breedt}, {Hollands}, {Izquierdo}  \& {Pelisoli}}{{Manser}
  et~al.}{2020}]{ManserEtAl-2020}
{Manser} C.~J.,  {G{\"a}nsicke} B.~T.,  {Gentile Fusillo} N.~P.,  {Ashley} R.,
  {Breedt} E.,  {Hollands} M.,  {Izquierdo} P.,   {Pelisoli} I.,  2020, \mn@doi
  [Monthly Notices of the Royal Astronomical Society] {10.1093/mnras/staa359},
  \href {https://ui.adsabs.harvard.edu/abs/2020MNRAS.493.2127M} {493, 2127}

\bibitem[\protect\citeauthoryear{{McKinnon} et~al.,}{{McKinnon}
  et~al.}{2020}]{McKinnonEtAl-2020}
{McKinnon} W.~B.,  et~al., 2020, \mn@doi [Science] {10.1126/science.aay6620},
  \href {https://ui.adsabs.harvard.edu/abs/2020Sci...367.6620M} {367, aay6620}

\bibitem[\protect\citeauthoryear{{Melis} \& {Dufour}}{{Melis} \&
  {Dufour}}{2017}]{MelisDufour-2017}
{Melis} C.,  {Dufour} P.,  2017, \mn@doi [The Astrophysical Journal]
  {10.3847/1538-4357/834/1/1}, \href
  {https://ui.adsabs.harvard.edu/abs/2017ApJ...834....1M} {834, 1}

\bibitem[\protect\citeauthoryear{{Melis}, {Jura}, {Albert}, {Klein}  \&
  {Zuckerman}}{{Melis} et~al.}{2010}]{MelisEtAl-2010}
{Melis} C.,  {Jura} M.,  {Albert} L.,  {Klein} B.,   {Zuckerman} B.,  2010,
  \mn@doi [The Astrophysical Journal] {10.1088/0004-637X/722/2/1078}, \href
  {https://ui.adsabs.harvard.edu/abs/2010ApJ...722.1078M} {722, 1078}

\bibitem[\protect\citeauthoryear{{Melis}, {Klein}, {Doyle}, {Weinberger},
  {Zuckerman}  \& {Dufour}}{{Melis} et~al.}{2020}]{MelisEtAl-2020}
{Melis} C.,  {Klein} B.,  {Doyle} A.~E.,  {Weinberger} A.~J.,  {Zuckerman} B.,
   {Dufour} P.,  2020, arXiv e-prints, \href
  {https://ui.adsabs.harvard.edu/abs/2020arXiv201003695M} {p. arXiv:2010.03695}

\bibitem[\protect\citeauthoryear{{Metzger}, {Rafikov}  \&
  {Bochkarev}}{{Metzger} et~al.}{2012}]{MetzgerEtAl-2012}
{Metzger} B.~D.,  {Rafikov} R.~R.,   {Bochkarev} K.~V.,  2012, \mn@doi [Monthly
  Notices of the Royal Astronomical Society]
  {10.1111/j.1365-2966.2012.20895.x}, \href
  {https://ui.adsabs.harvard.edu/abs/2012MNRAS.423..505M} {423, 505}

\bibitem[\protect\citeauthoryear{{Michaely} \& {Perets}}{{Michaely} \&
  {Perets}}{2014}]{MichaelyPerets-2014}
{Michaely} E.,  {Perets} H.~B.,  2014, \mn@doi [The Astrophysical Journal]
  {10.1088/0004-637X/794/2/122}, \href
  {http://adsabs.harvard.edu/abs/2014ApJ...794..122M} {794, 122}

\bibitem[\protect\citeauthoryear{{Mustill} \& {Villaver}}{{Mustill} \&
  {Villaver}}{2012}]{MustillVillaver-2012}
{Mustill} A.~J.,  {Villaver} E.,  2012, \mn@doi [The Astrophysical Journal]
  {10.1088/0004-637X/761/2/121}, \href
  {http://adsabs.harvard.edu/abs/2012ApJ...761..121M} {761, 121}

\bibitem[\protect\citeauthoryear{{Mustill}, {Veras}  \& {Villaver}}{{Mustill}
  et~al.}{2014}]{MustillEtAl-2014}
{Mustill} A.~J.,  {Veras} D.,   {Villaver} E.,  2014, \mn@doi [Monthly Notices
  of the Royal Astronomical Society] {10.1093/mnras/stt1973}, \href
  {http://adsabs.harvard.edu/abs/2014MNRAS.437.1404M} {437, 1404}

\bibitem[\protect\citeauthoryear{{Muto}, {Takeuchi}  \& {Ida}}{{Muto}
  et~al.}{2011}]{MutoEtAl-2011}
{Muto} T.,  {Takeuchi} T.,   {Ida} S.,  2011, \mn@doi [The Astrophysical
  Journal] {10.1088/0004-637X/737/1/37}, \href
  {https://ui.adsabs.harvard.edu/abs/2011ApJ...737...37M} {737, 37}

\bibitem[\protect\citeauthoryear{{Nixon}, {Pringle}, {Coughlin}, {Swan}  \&
  {Farihi}}{{Nixon} et~al.}{2020}]{NixonEtAl-2020}
{Nixon} C.~J.,  {Pringle} J.~E.,  {Coughlin} E.~R.,  {Swan} A.,   {Farihi} J.,
  2020, arXiv e-prints, \href
  {https://ui.adsabs.harvard.edu/abs/2020arXiv200607639N} {p. arXiv:2006.07639}

\bibitem[\protect\citeauthoryear{{O'Connor} \& {Lai}}{{O'Connor} \&
  {Lai}}{2020}]{OConnorLai-2020}
{O'Connor} C.~E.,  {Lai} D.,  2020, arXiv e-prints, \href
  {https://ui.adsabs.harvard.edu/abs/2020arXiv200505977O} {p. arXiv:2005.05977}

\bibitem[\protect\citeauthoryear{{O'Rourke} et~al.,}{{O'Rourke}
  et~al.}{2020}]{ORourkeEtAl-2020}
{O'Rourke} L.,  et~al., 2020, \mn@doi [Nature] {10.1038/s41586-020-2834-3},
  \href {https://ui.adsabs.harvard.edu/abs/2020Natur.586..697O} {586, 697}

\bibitem[\protect\citeauthoryear{{Ostriker}}{{Ostriker}}{1999}]{Ostriker-1999}
{Ostriker} E.~C.,  1999, \mn@doi [The Astrophysical Journal] {10.1086/306858},
  \href {https://ui.adsabs.harvard.edu/abs/1999ApJ...513..252O} {513, 252}

\bibitem[\protect\citeauthoryear{{Payne}, {Veras}, {Holman}  \&
  {G{\"a}nsicke}}{{Payne} et~al.}{2016}]{PayneEtAl-2016}
{Payne} M.~J.,  {Veras} D.,  {Holman} M.~J.,   {G{\"a}nsicke} B.~T.,  2016,
  \mn@doi [Monthly Notices of the Royal Astronomical Society]
  {10.1093/mnras/stv2966}, \href
  {http://adsabs.harvard.edu/abs/2016MNRAS.457..217P} {457, 217}

\bibitem[\protect\citeauthoryear{{Payne}, {Veras}, {G{\"a}nsicke}  \&
  {Holman}}{{Payne} et~al.}{2017}]{PayneEtAl-2017}
{Payne} M.~J.,  {Veras} D.,  {G{\"a}nsicke} B.~T.,   {Holman} M.~J.,  2017,
  \mn@doi [Monthly Notices of the Royal Astronomical Society]
  {10.1093/mnras/stw2585}, \href
  {http://adsabs.harvard.edu/abs/2017MNRAS.464.2557P} {464, 2557}

\bibitem[\protect\citeauthoryear{{Perets} \& {Kratter}}{{Perets} \&
  {Kratter}}{2012}]{PeretsKratter-2012}
{Perets} H.~B.,  {Kratter} K.~M.,  2012, \mn@doi [The Astrophysical Journal]
  {10.1088/0004-637X/760/2/99}, \href
  {http://adsabs.harvard.edu/abs/2012ApJ...760...99P} {760, 99}

\bibitem[\protect\citeauthoryear{{Petrovic}}{{Petrovic}}{2002}]{Petrovic2002}
{Petrovic} J.,  2002, in 34th COSPAR Scientific Assembly.

\bibitem[\protect\citeauthoryear{{Petrovich} \& {Mu{\~n}oz}}{{Petrovich} \&
  {Mu{\~n}oz}}{2017}]{PetrovichMunoz-2017}
{Petrovich} C.,  {Mu{\~n}oz} D.~J.,  2017, \mn@doi [Astrophysical Journal]
  {10.3847/1538-4357/834/2/116}, \href
  {http://adsabs.harvard.edu/abs/2017ApJ...834..116P} {834, 116}

\bibitem[\protect\citeauthoryear{{Podolak}, {Pollack}  \& {Reynolds}}{{Podolak}
  et~al.}{1988}]{Podolak1988}
{Podolak} M.,  {Pollack} J.~B.,   {Reynolds} R.~T.,  1988, \mn@doi [Icarus]
  {10.1016/0019-1035(88)90090-5}, \href
  {http://adsabs.harvard.edu/abs/1988Icar...73..163P} {73, 163}

\bibitem[\protect\citeauthoryear{{Popova}, {Borovi{\v c}ka}, {Hartmann},
  {Spurn{\'y}}, {Gnos}, {Nemtchinov}  \& {Trigo-Rodr{\'{\i}}guez}}{{Popova}
  et~al.}{2011}]{Popova2011}
{Popova} O.,  {Borovi{\v c}ka} J.,  {Hartmann} W.~K.,  {Spurn{\'y}} P.,  {Gnos}
  E.,  {Nemtchinov} I.,   {Trigo-Rodr{\'{\i}}guez} J.~M.,  2011, \mn@doi
  [Meteoritics and Planetary Science] {10.1111/j.1945-5100.2011.01247.x}, \href
  {http://adsabs.harvard.edu/abs/2011M%26PS...46.1525P} {46, 1525}

\bibitem[\protect\citeauthoryear{{Pravec}, {Harris}  \& {Michalowski}}{{Pravec}
  et~al.}{2002}]{PravecEtAl-2002}
{Pravec} P.,  {Harris} A.~W.,   {Michalowski} T.,  2002, {Asteroid Rotations}.
University of Arizona Press, Tucson, p.113-122, pp 113--122

\bibitem[\protect\citeauthoryear{{Rafikov}}{{Rafikov}}{2011}]{Rafikov-2011}
{Rafikov} R.~R.,  2011, \mn@doi [Monthly Notices of the Royal Astronomical
  Society: Letters] {10.1111/j.1745-3933.2011.01096.x}, \href
  {http://adsabs.harvard.edu/abs/2011MNRAS.416L..55R} {416, L55}

\bibitem[\protect\citeauthoryear{{Rafikov}}{{Rafikov}}{2018}]{Rafikov-2018}
{Rafikov} R.~R.,  2018, \mn@doi [The Astrophysical Journal]
  {10.3847/1538-4357/aac5ef}, \href
  {https://ui.adsabs.harvard.edu/abs/2018ApJ...861...35R} {861, 35}

\bibitem[\protect\citeauthoryear{{Rappaport}, {Gary}, {Vanderburg}, {Xu},
  {Pooley}  \& {Mukai}}{{Rappaport} et~al.}{2018}]{RappaportEtAl-2018}
{Rappaport} S.,  {Gary} B.~L.,  {Vanderburg} A.,  {Xu} S.,  {Pooley} D.,
  {Mukai} K.,  2018, \mn@doi [Monthly Notices of the Royal Astronomical Societ]
  {10.1093/mnras/stx2663}, \href
  {https://ui.adsabs.harvard.edu/abs/2018MNRAS.474..933R} {474, 933}

\bibitem[\protect\citeauthoryear{{Rebassa-Mansergas}, {Solano}, {Xu},
  {Rodrigo}, {Jim{\'e}nez-Esteban}  \& {Torres}}{{Rebassa-Mansergas}
  et~al.}{2019}]{RebassaMansergasEtAl-2019}
{Rebassa-Mansergas} A.,  {Solano} E.,  {Xu} S.,  {Rodrigo} C.,
  {Jim{\'e}nez-Esteban} F.~M.,   {Torres} S.,  2019, \mn@doi [Monthly Notices
  of the Royal Astronomical Society] {10.1093/mnras/stz2423}, \href
  {https://ui.adsabs.harvard.edu/abs/2019MNRAS.489.3990R} {489, 3990}

\bibitem[\protect\citeauthoryear{{Rein} \& {Liu}}{{Rein} \&
  {Liu}}{2012}]{ReboundMain}
{Rein} H.,  {Liu} S.-F.,  2012, \mn@doi [Astronomy \& Astrophysics]
  {10.1051/0004-6361/201118085}, \href
  {http://adsabs.harvard.edu/abs/2012A%26A...537A.128R} {537, A128}

\bibitem[\protect\citeauthoryear{{Rein} \& {Spiegel}}{{Rein} \&
  {Spiegel}}{2015}]{ReboundIAS15}
{Rein} H.,  {Spiegel} D.~S.,  2015, \mn@doi [Monthly Notices of the Royal
  Astronomical Society] {10.1093/mnras/stu2164}, \href
  {http://adsabs.harvard.edu/abs/2015MNRAS.446.1424R} {446, 1424}

\bibitem[\protect\citeauthoryear{{Rephaeli} \& {Salpeter}}{{Rephaeli} \&
  {Salpeter}}{1980}]{RaphaeliEtAl-1980}
{Rephaeli} Y.,  {Salpeter} E.~E.,  1980, \mn@doi [The Astrophysical Journal]
  {10.1086/158202}, \href
  {https://ui.adsabs.harvard.edu/abs/1980ApJ...240...20R} {240, 20}

\bibitem[\protect\citeauthoryear{{Rocchetto}, {Farihi}, {G{\"a}nsicke}  \&
  {Bergfors}}{{Rocchetto} et~al.}{2015}]{RocchettoEtAl-2015}
{Rocchetto} M.,  {Farihi} J.,  {G{\"a}nsicke} B.~T.,   {Bergfors} C.,  2015,
  \mn@doi [Monthly Notices of the Royal Astronomical Society]
  {10.1093/mnras/stv282}, \href
  {https://ui.adsabs.harvard.edu/abs/2015MNRAS.449..574R} {449, 574}

\bibitem[\protect\citeauthoryear{{Rogers}, {Xu}, {Bonsor}, {Hodgkin}, {Su},
  {von Hippel}  \& {Jura}}{{Rogers} et~al.}{2020}]{RogersEtAl-2020}
{Rogers} L.~K.,  {Xu} S.,  {Bonsor} A.,  {Hodgkin} S.,  {Su} K. Y.~L.,  {von
  Hippel} T.,   {Jura} M.,  2020, \mn@doi [Monthly Notices of the Royal
  Astronomical Society] {10.1093/mnras/staa873}, \href
  {https://ui.adsabs.harvard.edu/abs/2020MNRAS.494.2861R} {494, 2861}

\bibitem[\protect\citeauthoryear{{Rozner}, {Veras}  \& {Perets}}{{Rozner}
  et~al.}{2020a}]{RoznerEtAl-2020b}
{Rozner} M.,  {Veras} D.,   {Perets} H.~B.,  2020a, arXiv e-prints, p.
  arXiv:2011.12299

\bibitem[\protect\citeauthoryear{{Rozner}, {Grishin}  \& {Perets}}{{Rozner}
  et~al.}{2020b}]{RoznerEtAl-2020a}
{Rozner} M.,  {Grishin} E.,   {Perets} H.~B.,  2020b, \mn@doi [Monthly Notices
  of the Royal Astronomical Society] {10.1093/mnras/staa1864}, \href
  {https://ui.adsabs.harvard.edu/abs/2020MNRAS.496.4827R} {496, 4827}

\bibitem[\protect\citeauthoryear{{Shappee} \& {Thompson}}{{Shappee} \&
  {Thompson}}{2013}]{ShapeeThompson-2013}
{Shappee} B.~J.,  {Thompson} T.~A.,  2013, \mn@doi [The Astrophysical Journal]
  {10.1088/0004-637X/766/1/64}, \href
  {http://adsabs.harvard.edu/abs/2013ApJ...766...64S} {766, 64}

\bibitem[\protect\citeauthoryear{{Smallwood}, {Martin}, {Livio}  \&
  {Lubow}}{{Smallwood} et~al.}{2018}]{SmallwoodEtAl-2018}
{Smallwood} J.~L.,  {Martin} R.~G.,  {Livio} M.,   {Lubow} S.~H.,  2018,
  \mn@doi [Monthly Notices of the Royal Astronomical Society]
  {10.1093/mnras/sty1819}, \href
  {https://ui.adsabs.harvard.edu/abs/2018MNRAS.480...57S} {480, 57}

\bibitem[\protect\citeauthoryear{{Smrekar}, {Cintala}  \& {Horz}}{{Smrekar}
  et~al.}{1985}]{SmrekarEtAl-1985}
{Smrekar} S.,  {Cintala} M.~J.,   {Horz} F.,  1985, in Lunar and Planetary
  Science Conference. Lunar and Planetary Science Conference.
pp 793--794

\bibitem[\protect\citeauthoryear{{Spohn} et~al.,}{{Spohn}
  et~al.}{2015}]{SpohnEtAl-2015}
{Spohn} T.,  et~al., 2015, \mn@doi [Science] {10.1126/science.aab0464}, \href
  {https://ui.adsabs.harvard.edu/abs/2015Sci...349b0464S} {349, 2.464}

\bibitem[\protect\citeauthoryear{{Stephan}, {Naoz}  \& {Zuckerman}}{{Stephan}
  et~al.}{2017}]{StephanEtAl-2017}
{Stephan} A.~P.,  {Naoz} S.,   {Zuckerman} B.,  2017, \mn@doi [The
  Astrophysical Journal Letters] {10.3847/2041-8213/aa7cf3}, \href
  {http://adsabs.harvard.edu/abs/2017ApJ...844L..16S} {844, L16}

\bibitem[\protect\citeauthoryear{{Stone}, {Metzger}  \& {Loeb}}{{Stone}
  et~al.}{2015}]{StoneEtAl-2015}
{Stone} N.,  {Metzger} B.~D.,   {Loeb} A.,  2015, \mn@doi [Monthly Notices of
  the Royal Astronomical Society] {10.1093/mnras/stu2718}, \href
  {http://adsabs.harvard.edu/abs/2015MNRAS.448..188S} {448, 188}

\bibitem[\protect\citeauthoryear{{Swan}, {Farihi}  \& {Wilson}}{{Swan}
  et~al.}{2019}]{SwanEtAl-2019}
{Swan} A.,  {Farihi} J.,   {Wilson} T.~G.,  2019, \mn@doi [Monthly Notices of
  the Royal Astronomical Society Letters] {10.1093/mnrasl/slz014}, \href
  {https://ui.adsabs.harvard.edu/abs/2019MNRAS.484L.109S} {484, L109}

\bibitem[\protect\citeauthoryear{{Swan}, {Farihi}, {Wilson}  \&
  {Parsons}}{{Swan} et~al.}{2020}]{SwanEtAl-2020}
{Swan} A.,  {Farihi} J.,  {Wilson} T.~G.,   {Parsons} S.~G.,  2020, \mn@doi
  [Monthly Notices of the Royal Astronomical Society] {10.1093/mnras/staa1688},
  \href {https://ui.adsabs.harvard.edu/abs/2020MNRAS.496.5233S} {496, 5233}

\bibitem[\protect\citeauthoryear{{Tielens}, {McKee}, {Seab}  \&
  {Hollenbach}}{{Tielens} et~al.}{1994}]{TielensEtAl-1994}
{Tielens} A.~G.~G.~M.,  {McKee} C.~F.,  {Seab} C.~G.,   {Hollenbach} D.~J.,
  1994, \mn@doi [The Astrophysical Journal] {10.1086/174488}, \href
  {https://ui.adsabs.harvard.edu/abs/1994ApJ...431..321T} {431, 321}

\bibitem[\protect\citeauthoryear{{Vanderburg} et~al.,}{{Vanderburg}
  et~al.}{2015}]{VanderburgEtAl-2015}
{Vanderburg} A.,  et~al., 2015, \mn@doi [Nature] {10.1038/nature15527}, \href
  {http://adsabs.harvard.edu/abs/2015Natur.526..546V} {526, 546}

\bibitem[\protect\citeauthoryear{{Veras}}{{Veras}}{2016}]{Veras-2016}
{Veras} D.,  2016, \mn@doi [Royal Society Open Science] {10.1098/rsos.150571},
  \href {http://adsabs.harvard.edu/abs/2016RSOS....3.0571V} {3, 150571}

\bibitem[\protect\citeauthoryear{{Veras} \& {G{\"a}nsicke}}{{Veras} \&
  {G{\"a}nsicke}}{2015}]{VerasGansicke-2015}
{Veras} D.,  {G{\"a}nsicke} B.~T.,  2015, \mn@doi [Monthly Notices of the Royal
  Astronomical Society] {10.1093/mnras/stu2475}, \href
  {http://adsabs.harvard.edu/abs/2015MNRAS.447.1049V} {447, 1049}

\bibitem[\protect\citeauthoryear{{Veras}, {Leinhardt}, {Bonsor}  \&
  {G{\"a}nsicke}}{{Veras} et~al.}{2014}]{VerasEtAl-2014}
{Veras} D.,  {Leinhardt} Z.~M.,  {Bonsor} A.,   {G{\"a}nsicke} B.~T.,  2014,
  \mn@doi [Monthly Notices of the Royal Astronomical Society]
  {10.1093/mnras/stu1871}, \href
  {http://adsabs.harvard.edu/abs/2014MNRAS.445.2244V} {445, 2244}

\bibitem[\protect\citeauthoryear{{Veras}, {Leinhardt}, {Eggl}  \&
  {G{\"a}nsicke}}{{Veras} et~al.}{2015}]{VerasEtAl-2015}
{Veras} D.,  {Leinhardt} Z.~M.,  {Eggl} S.,   {G{\"a}nsicke} B.~T.,  2015,
  \mn@doi [Monthly Notices of the Royal Astronomical Society]
  {10.1093/mnras/stv1195}, \href
  {http://adsabs.harvard.edu/abs/2015MNRAS.451.3453V} {451, 3453}

\bibitem[\protect\citeauthoryear{{Wang} et~al.,}{{Wang}
  et~al.}{2019}]{WangEtAl-2019}
{Wang} T.-g.,  et~al., 2019, \mn@doi [The Astrophysical Journal Letters]
  {10.3847/2041-8213/ab53ed}, \href
  {https://ui.adsabs.harvard.edu/abs/2019ApJ...886L...5W} {886, L5}

\bibitem[\protect\citeauthoryear{{Weidenschilling}}{{Weidenschilling}}{1977}]{wei77}
{Weidenschilling} S.~J.,  1977, \mn@doi [Astrophysics and Space Science]
  {10.1007/BF00642464}, \href
  {https://ui.adsabs.harvard.edu/abs/1977Ap&SS..51..153W} {51, 153}

\bibitem[\protect\citeauthoryear{{Wilson}, {G{\"a}nsicke}, {Koester}, {Raddi},
  {Breedt}, {Southworth}  \& {Parsons}}{{Wilson}
  et~al.}{2014}]{WilsonEtAl-2014}
{Wilson} D.~J.,  {G{\"a}nsicke} B.~T.,  {Koester} D.,  {Raddi} R.,  {Breedt}
  E.,  {Southworth} J.,   {Parsons} S.~G.,  2014, \mn@doi [Monthly Notices of
  the Royal Astronomical Society] {10.1093/mnras/stu1876}, \href
  {https://ui.adsabs.harvard.edu/abs/2014MNRAS.445.1878W} {445, 1878}

\bibitem[\protect\citeauthoryear{{Wyatt}, {Clarke}  \& {Booth}}{{Wyatt}
  et~al.}{2011}]{WyattEtAl-2011}
{Wyatt} M.~C.,  {Clarke} C.~J.,   {Booth} M.,  2011, \mn@doi [Celestial
  Mechanics and Dynamical Astronomy] {10.1007/s10569-011-9345-3}, \href
  {https://ui.adsabs.harvard.edu/abs/2011CeMDA.111....1W} {111, 1}

\bibitem[\protect\citeauthoryear{{Xu} \& {Jura}}{{Xu} \&
  {Jura}}{2012}]{XuJura-2012}
{Xu} S.,  {Jura} M.,  2012, \mn@doi [The Astrophysical Journal]
  {10.1088/0004-637X/745/1/88}, \href
  {http://adsabs.harvard.edu/abs/2012ApJ...745...88X} {745, 88}

\bibitem[\protect\citeauthoryear{{Xu} \& {Jura}}{{Xu} \&
  {Jura}}{2014}]{XuJura-2014}
{Xu} S.,  {Jura} M.,  2014, \mn@doi [The Astrophysical Journal Letters]
  {10.1088/2041-8205/792/2/L39}, \href
  {https://ui.adsabs.harvard.edu/abs/2014ApJ...792L..39X} {792, L39}

\bibitem[\protect\citeauthoryear{{Xu} et~al.,}{{Xu} et~al.}{2018}]{XuEtAl-2018}
{Xu} S.,  et~al., 2018, \mn@doi [The Astrophysical Journal]
  {10.3847/1538-4357/aadcfe}, \href
  {https://ui.adsabs.harvard.edu/abs/2018ApJ...866..108X} {866, 108}

\bibitem[\protect\citeauthoryear{{Xu} et~al.,}{{Xu} et~al.}{2019}]{XuEtAl-2019}
{Xu} S.,  et~al., 2019, \mn@doi [The Astronomical Journal]
  {10.3847/1538-3881/ab1b36}, \href
  {https://ui.adsabs.harvard.edu/abs/2019AJ....157..255X} {157, 255}

\bibitem[\protect\citeauthoryear{{Zhou} et~al.,}{{Zhou}
  et~al.}{2016}]{ZhouEtAl-2016}
{Zhou} G.,  et~al., 2016, \mn@doi [Monthly Notices of the Royal Astronomical
  Society] {10.1093/mnras/stw2286}, \href
  {https://ui.adsabs.harvard.edu/abs/2016MNRAS.463.4422Z} {463, 4422}

\bibitem[\protect\citeauthoryear{{Zuckerman}, {Koester}, {Reid}  \&
  {H{\"u}nsch}}{{Zuckerman} et~al.}{2003}]{ZuckermanEtAl-2003}
{Zuckerman} B.,  {Koester} D.,  {Reid} I.~N.,   {H{\"u}nsch} M.,  2003, \mn@doi
  [The Astrophysical Journal,] {10.1086/377492}, \href
  {http://adsabs.harvard.edu/abs/2003ApJ...596..477Z} {596, 477}

\bibitem[\protect\citeauthoryear{{Zuckerman}, {Melis}, {Klein}, {Koester}  \&
  {Jura}}{{Zuckerman} et~al.}{2010}]{ZuckermanEtAl-2010}
{Zuckerman} B.,  {Melis} C.,  {Klein} B.,  {Koester} D.,   {Jura} M.,  2010,
  \mn@doi [The Astrophysical Journal] {10.1088/0004-637X/722/1/725}, \href
  {http://adsabs.harvard.edu/abs/2010ApJ...722..725Z} {722, 725}

\bibitem[\protect\citeauthoryear{{van Lieshout}, {Kral}, {Charnoz}, {Wyatt}  \&
  {Shannon}}{{van Lieshout} et~al.}{2018}]{vanLieshoutEtAl-2018}
{van Lieshout} R.,  {Kral} Q.,  {Charnoz} S.,  {Wyatt} M.~C.,   {Shannon} A.,
  2018, \mn@doi [Monthly Notices of the Royal Astronomical Society]
  {10.1093/mnras/sty1271}, \href
  {https://ui.adsabs.harvard.edu/abs/2018MNRAS.480.2784V} {480, 2784}

\makeatother
\end{thebibliography}

	
	\appendix
	
	\newpage
	
	\section{Two vs. one intersection points}\label{A:IntersectionPoints}
	In Section \ref{SS:analytic} we point out that energy loss in Equation \ref{eq:de} ensues from a single intersection between a fragment and the compact disc, which in turn originates from our simplification that the argument of pericentre is assumed zero.
	
	As already mentioned in Section \ref{SS:FragDicIntersection}, the general case may result in two intersection points with the compact disc instead of one. For example, let us consider the orthogonal case where the argument of pericentre equals $\pi/2$. Then it is easy to show that the true anomaly at the two crossing points is $\pi/2$, and the distance between the central star and the two crossing points is $2q$ instead of $q$. Now having two intersection points, $2\Delta E (2q)$ replaces $\Delta E (q)$ in Equation \ref{eq:de}. $v_{rel}^2$ goes like $\propto 1/r$ for large eccentricities, so increasing $q$ by a factor of 2 cancels out with the 2 prefactor. The only difference then comes from replacing $\Sigma(q)$ with $\Sigma(2q)$, which translates into a factor $2^{-\beta}$, or $\sim$0.35. This difference is merely of order unity, and can therefore be neglected.

	\section{Gap filling timescale}\label{A:GapFillingTimescale}
	\subsection{Estimating the diffusion filling time}
	Consider a disc of total mass $M$ with an outer boundary $r_{\rm out}.$ For simplicity, the radial profile of the compact disc surface density is now neglected, and $r_{\rm in}=0$. The scale height is generally given by $h=fr_{{\rm out}},$ where $f$ is a small number which depends on the disc properties and will be discussed later. The latter dictates a relationship between the velocity dispersion in the disc vertical axis and the Keplerian velocity $v_{\rm z}/v_{\rm K}\sim h/r_{\rm out}\sim f$, noting also that $v_{\rm h}=2v_{\rm z}$ is the horizontal velocity dispersion.
	
	Consider that a cylindrical gap of radius $R$ (i.e comparable to the radius of the interacting fragment) has been carved in the compact disc. The dusty particles around it will scatter to fill it. A particle at location $D$ from the gap moving horizontally at a random direction with velocity $v_{\rm h}$. After passing a mean free path $\lambda$ it encounters and scatters again at a different horizontal velocity and preforms a random walk. From diffusion, the number of steps for a particle to traverse a distance $D$ and fill the gap is $X=(D/\lambda)^{2}$. The mean time between collisions is $\tau_{{\rm coll}}=\lambda/v_{\rm h}$ thus the total diffusion time to fill the gap is 
	
	\begin{equation} \label{eq:tfill} \tau_{{\rm fill}}=X \tau_{{\rm coll}}=\frac{D^{2}}{\lambda^{2}}\frac{\lambda}{v_{\rm h}}=\frac{D^{2}}{\lambda v_{\rm h}}=\frac{D^{2}}{2 \lambda f v_{\rm K}}.
	\end{equation}
	
	It is reasonable to assume that in order to fill a gap of dimension $R$ the particle must also travel a similar distance, so we take $D\approx R$.
	
	\subsection{Estimation of the mean free path}
	Generally, the mean free path is $\lambda=1/nA$ where $A=\pi r_{\rm d}^2$ is the surface area given a dust grain size $r_{\rm d}$, and the dust number density $n$ is given by
	
	\begin{equation}
		n=\frac{\rho_{\rm disc}}{m_{d}}=\frac{3 \rho_{\rm disc}}{4\pi \rho_{\rm d} r_{\rm d}^{3}},
	\end{equation}
	
	\noindent where $\rho_{\rm disc}$ is the density of a the pre-existing compact disc, $m_{d}$ the mass of a dust grain and $\rho_{d}$ the density of a dust grain. The mean free path is then
	
	\begin{equation} \label{eq:mfp1}
		\lambda=\frac{4\pi\rho_{\rm d} r_{\rm d}^{3}}{3 \rho_{{\rm disc}}\pi r_{\rm d}^{2}}=\frac{4}{3}\frac{\rho_{\rm d}}{\rho_{\rm disc}}r_{\rm d}=\frac{4}{3}f\frac{\rho_{\rm d} r_{\rm out}^3}{M}r_{\rm d}.
	\end{equation}
	
	The diffusion filling time is obtained from Equation \ref{eq:tfill} as
	\begin{equation} \label{eq:tdif2}
		\tau_{{\rm fill}}=\frac{3M R^2}{8f^2 \rho_{\rm d} r_{\rm out}^3 r_{\rm d} v_{\rm K}}=\frac{3M R^2 q^{1/2}}{8h^2 \rho_{\rm d} r_{\rm out} r_{\rm d} (GM_{\rm WD})^{1/2}}.
	\end{equation}
	
	Note that $\tau_{\rm fill} \propto M$, meaning that as the compact disc mass decreases, it takes less time to fill the gap. This might at first seem counter-intuitive, but in fact it is a logical outcome. When $M$ lowers, the disc density $\rho_{\rm disc}$ lowers. Then the dust number density decreases, so that the mean free path is larger. Although the time between collisions $\tau_{\rm coll}$ increases, $X$ decreases more and less steps are required to traverse the same distance.
	
	\subsection{Comparison with existing literature}
	Above we derive the filling timescale from first principles. In order to verify our result we compare it with previous calculation of gap filling by \cite{BromleyKenyon-2013} applied to Saturn's rings. There the time to fill the gap is 

	\begin{equation} \label{eq:tfillBromely}
		\tau_{{\rm fill}}=\frac{R^{2}}{\nu_{\rm rad}},
	\end{equation}

	\noindent where the radial viscosity $\nu_{\rm rad}$ is given by \citep{GoldreichTremaine-1978}:
	
	\begin{equation} \label{eq:radialViscosity}
		\nu_{\rm rad}=\frac{v_{\rm h}^2P}{4\pi}\frac{\tau_{\rm d}}{1+\tau_{\rm d}^2},
	\end{equation}

	\noindent such that $P$ is the orbital period and $\tau_d=3\Sigma/4\rho_{\rm d} r_{\rm d}$ is the "collisional depth" of the dust, analogous to optical depth in the context of photon diffusion\footnote{The optical depth concept is borrowed from photon diffusion, and estimates the number of collisions a dust particle need to traverse while passing in the disc. To avoid confusion, we call it "collisional depth" to better relate to the nature of the interactions.}.
	
	In the optically thick limit $\tau_{\rm d} \gg 1$, Equation \ref{eq:radialViscosity} reduces to $\nu_{\rm rad}=v_{\rm h}^2P/4\pi \tau_{\rm d}$. In terms of the disc parameters, we recast $P=2\pi r_{\rm out}/v_{\rm K}=\pi r_{\rm out} fv_{\rm h}$ and $\Sigma \approx M/r_{\rm out}^2$. We then replace $\nu_{\rm rad}$ in Equation \ref{eq:tfillBromely} and obtain the exact same filling time as in Equation \ref{eq:tdif2}.
		
	Indeed by taking typical values for $\rho_{\rm d}$, $r_{\rm d}$ and $r_{\rm out}$, we have $\tau_d \gg 1$ for all but the least massive discs and the gap filling is by horizontal diffusion. If $\tau_d \approx 1$ then the gap filling is not diffusive and therefore even quicker. We conclude that Equation \ref{eq:tdif2} is correct and may be regarded as an upper-limit timescale.

	\section{Gravitational instabilities}\label{A:GravInstabilities}	
	Self gravity is important when the Toomre $Q$ is close to $1$. For gaseous discs, the Toomre $Q$ parameter is given by
	
	\begin{equation}\label{eq:ToomreParam}
		Q=\frac{c_{s}\Omega}{\pi G\Sigma}=\frac{c_{s}^{2}}{\pi G\Sigma h}\approx \frac{c_{s}^2 r_{\rm out}^2}{\pi G M h},
	\end{equation}
	
	\noindent where the total compact disc mass is $M\sim\Sigma r_{{\rm out}}^{2}$. Taking plausible values $c_{s}\sim1{\rm km/s}$ and $h\sim1000$ km, the compact disc mass $M$ has to be larger than $\sim10^{29}$ g, which is much more massive than any disc in our study. 
	
	For dusty debris discs, the sound speed is replaced by the appropriate dispersion velocity. The smallest dispersion velocity is the vertical one, $v_{\rm z}$, which minimizes $Q$. It is comparable to the aforementioned sound speed for scale height of $h\sim 10^3\ \rm km$, since $v_{\rm z} \cong v_{\rm K} f \approx v_{\rm K} h / r_{\rm out}$, or $\sqrt{GM_{\rm WD}} h / r_{\rm out}^{3/2}$.
	
	Q$\approx$1 from Equation \ref{eq:ToomreParam} then yields the limit for the scale height as
	
	\begin{equation}\label{eq:GravStabCond}
		h=\frac{\pi r_{\rm out} M}{M_{\rm WD}} \cong 1~\rm{m}.
	\end{equation}
	
	\noindent where we take $M=10^{24}$ g, the largest mass in our parameter space and $M_{\rm WD}=0.6M_\odot$, a typical WD mass. The conclusion is that close to the Roche radius, the discs are well stable unless $h$ is smaller than about a meter, or much less if we take a smaller $M$.

	\section{Effective shattering yield}\label{A:ShatteringYield}
	The shattering yield, first discussed in Section \ref{SS:GasProduction}, is defined as the excavated (fragment) dust mass over the impactor dust mass. The study of \cite{KoschnyGrun-2001} provides an estimate for the shattering yield $Y_{shat}$ for pure dust by the following formula:
	
	\begin{equation}
		Y = 10^{-9} \rho_{\rm d} 2^{-b} m_d^{b-1} v_{\rm rel}^{2b} \label{eq:ShatteringYield},
	\end{equation}
	
	\noindent where $\rho_d$ and $m_d$ are the dust grain density and mass. The parameter $b$ depends on the material fraction of silicate versus ice. We adopt $b \cong 1.5$ for pure silicate \citep{SmrekarEtAl-1985}. Typical relative velocities $v_{\rm rel}$ from Section \ref{SS:GasProduction} are of the order of a few $\sim 10^2$ m $\times$ s$^{-1}$, leading to a shattering yield of approximately $Y=\sim 3000$ for micron-sized particles.
	
	However, the effective yield is lower than this value. The excavated material forms a dust cloud shielding the fragment surface. This ejecta in turn encounters more compact disc dust grains. At these velocities grain-grain collisions result in complete vaporization. In the event of full shielding, no other compact disc dust grain is able to impact the fragment's surface until the ejecta is fully vaporized, and only then a subsequent grain can reach the surface, giving an actual effective yield of $Y_{\rm eff}=1$. We will demonstrate this quantitatively.	
	
	In the fragment's frame of reference, consider the projectile compact disc dust grains with number density $n_1$ and cross section $A=\pi r_d^2$ hitting a fragment of size $R\gg r_d$ at velocity $v_1=v_{\rm rel}$. Each impact liberates $Y$ dust particles of the same size (for simplicity) from the fragment with velocity $v_2$. $Y$ is the yield, which is taken roughly as $3000$ (initially). From energy balance, if all the kinetic energy $0.5m_dv_1^2$ went into the kinetic energy of the fragments, $0.5Y m_d v_2^2$, we get $v_2=v_1/ \sqrt{Y}$.
	
	Consider now a swarm of ejecta particles formed at time $dt$ with total mass $dm_{\rm ejecta}=\rho_{\rm ejecta}Av_2 dt$. It was formed by compact disc projectile dust grains of mass $dm_{\rm disc} = \rho_{\rm disc} A v_1 dt$. The mass ratio is exactly the initial yield:
	
	\begin{equation}\label{eq:InitialYield}
		Y=\frac{dm_{\rm ejecta}}{dm_{\rm disc}}=\frac{\rho_{\rm ejecta}}{\rho_{\rm disc}}\frac{v_2}{v_1}=\frac{\rho_{\rm ejecta}}{\sqrt{Y}\rho_{\rm disc}}.
	\end{equation}
	
	If both the compact disc particles and the ejecta particles are "optically thick" such that the mean free path is very small, we have from Equation \ref{eq:mfp1}, $\lambda_{\rm disc} \sim (\rho_{\rm d}/\rho_{\rm disc} )r_{\rm d}$ for the disc particles and
	
	\begin{equation}\label{eq:MFPejecta}
		\lambda_{\rm ejecta} \sim \frac{\rho_{\rm d}}{\rho_{\rm ejecta}} r_{\rm d}=\lambda_{\rm disc}  \frac{\rho_{\rm disc}}{\rho_{\rm ejecta}}=\frac{\lambda_{\rm disc}}{Y^{3/2}},
	\end{equation} 
	
	\noindent where the last substitution came from rearranging Equation \ref{eq:InitialYield}.
	
	In terms of the compact disc mass $M_{\rm disc}$ and the scale height $h$, from Equation \ref{eq:mfp1} we obtain
	
	\begin{equation}\label{eq:MFPdisc}
		\lambda_{\rm disc} \approx \left( \frac{10^{22}~\rm{g}}{M_{\rm disc}} \right) \left( \frac{h}{10^3~\rm{km}} \right)~\rm{cm}
	\end{equation}
	
	Finally, the timescale for the ejecta particles to become "optically thick" is
	
	\begin{equation}\label{eq:MFPdisc}
		t \approx \frac{\lambda_{\rm ejecta}}{v_2} = \frac{\lambda_{\rm disc}}{Yv_1} \approx \left( \frac{10^{16}~\rm{g}}{M_{\rm disc}} \right) \left( \frac{h}{10^3~\rm{km}} \right) 10^{-6}~\rm{s}
	\end{equation}
	
	For a fragment crossing the compact disc face-on, the flight time through the disc is simply $t_{\rm flight}=h/v_{\rm rel}$. For typical fragment velocities, an optically thick ejecta blanket would form within a fraction of $t/t_{\rm flight}$ of the flight time, i.e. 1 millionth or less of the flight time.			
	
	If an optically thick cloud is present, as was just established, then only a small fraction of the projectile compact disc dust grains may hit the fragment and the effective yield $Y_{\rm eff}$ will then be much lower than 3000. To estimate $Y_{\rm eff}$, we want the number of ejecta grains to be in steady state, namely the generation rate $\mathcal{R}_{\rm gen}$ must be balanced by the encounter, or vaporization rate $\mathcal{R}_{\rm evap}$. For the generation rate, it is just the nominal hitting rate times the effective yield, $\mathcal{R}_{\rm gen}=Y_{\rm eff} n_1 A v_1$, while for the vaporization rate, in the rest frame of the ejecta swarm, it is $\mathcal{R}_{\rm evap}=n_1A(v_1+v_2)$, since the relative velocity between the grains is $v_1+v_2$. Comparing the rates and plugging in the velocities, we get  that the effective yield is $Y_{\rm eff}=(v_1+v_2)/v_1=1+\sqrt{1/Y}\approx 1$. The effective yield cannot exceed much more than unity if a thick could is present.

	\bsp	
	\label{lastpage}
\end{document}